\PassOptionsToPackage{warn}{textcomp}
\documentclass[letterpaper, 12pt]{article}
\usepackage{graphicx}
\usepackage{amsmath}
\usepackage{amsfonts}
\usepackage{amssymb}
\usepackage{amsthm}
\usepackage{bm}
\usepackage{accents}
\usepackage{statex}
\usepackage{mathptmx}
\usepackage{marginnote}
\usepackage{enumitem}
\usepackage{rotating}
\usepackage{fancyvrb}
\usepackage{epstopdf}
\usepackage{float}
\usepackage{cases}
\usepackage[doublespacing]{setspace}
\usepackage{natbib}
\usepackage{multirow}
\usepackage{dcolumn}
\usepackage{booktabs}
\usepackage{caption}
\usepackage{threeparttable}
\usepackage{longtable}
\usepackage{tabularx}
\usepackage{makecell}
\usepackage{siunitx}
\usepackage{geometry}
\usepackage{hyperref}
\usepackage{kbordermatrix}
\usepackage{subcaption}
\usepackage{pdfpages}
\usepackage{lscape}
\usepackage{url}
\usepackage{titling}
\usepackage{titlesec}
\usepackage{times}
\usepackage{comment}
\usepackage{datetime}
\usepackage{textcomp}
\usepackage{pdfpages}
\usepackage[symbol]{footmisc}
\usepackage[svgnames]{xcolor}
\hypersetup{colorlinks, citecolor=NavyBlue, filecolor=NavyBlue, linkcolor=NavyBlue, urlcolor=NavyBlue}
\makeatletter
\def\wideubar{\underaccent{{\cc@style\underline{\mskip8mu}}}}
\def\@pdfborder{0 0 0}
\newtheorem{hyp}{Hypothesis}

\numberwithin{corollary}{section}
\numberwithin{lemma}{section}
\numberwithin{proposition}{section}
\numberwithin{theorem}{section}
\numberwithin{remark}{section}
\titlelabel{\thetitle.\quad}
\titleformat{\subsection}{\bf\large\itshape}{\thesubsection.}{1em}{}
\newdateformat{monthyeardate}{%
  \monthname[\THEMONTH] \THEYEAR}
  
\renewcommand*{\thefootnote}{\fnsymbol{footnote}}
\graphicspath{{./Figures/}}
\begin{document}

\setlength{\droptitle}{-5em}

\title{\textbf{Joint News, Attention Spillover,\\ and Stock Returns}\thanks{%We are grateful to Stefano Giglio (Editor) and two anonymous referees for very insightful and helpful comments that improve the paper substantially. 
\footnotesize  
Previous versions of this paper have been circulated under the titles ``Media Network Based Investors' Attention: A Powerful Predictor of Market Premium" and ``News Co-occurrence, Attention Spillover and Return Predictability."  Li Guo, School of Economics, Fudan University and Shanghai Institute of International Finance and Economics; Lin Peng, Zicklin School of Business, Baruch College, the City University of New York; Yubo Tao, Faculty of Social Sciences and Faculty of Business Administration, University of Macau; Jun Tu, SJTU-BOC Institute of Technology and Finance, Antai College of Economics and Management, Shanghai Jiao Tong University, and Lee Kong Chian School of Business, Singapore Management University. Send correspondence to Jun Tu, tujun2022@sjtu.edu.cn (tujun@smu.edu.sg). We thank Zhi Da, Joseph Engelberg (discussant), Gang Li (discussant), Tse-Chun Lin (discussant), Oliver Linton, Jorida Papakroni (discussant), Nancy Xu (discussant), Dacheng Xiu, Guofu Zhou, and participants at the Asian Financial Association Annual Conference, the China International Conference in Finance, the Financial Management Association Annual Conference, the North American Summer Meeting of the Econometric Society, the Shanghai Financial Forefront Symposium, and the Singapore Management University Finance Summer Camp for their very helpful comments. Li Guo acknowledges financial support from the National Natural Science Foundation of China (Project No.~72003040). Lin Peng acknowledges the Krell Research Grant for financial support. Tu acknowledges financial support from the National Natural Science Foundation of China (No.~72342023) and the Singapore Management University through a research grant (Grant No. 22-LKCSB-SMU-018) from the Ministry of Education Academic Research Fund Tier 1. Yubo Tao acknowledges the financial support from the National Natural Science Foundation of China (Project No.~72303003) and the Startup Research Grant (No.~SRG2022-00016-FSS) from the University of Macau. All authors contributed equally to this paper. }}

\author{ Li Guo
\hspace{10mm} Lin Peng
\hspace{10mm} Yubo Tao
\hspace{10mm} Jun Tu }

\date{}

\begin{titlingpage}
\maketitle

\vspace{-2cm}

\begin{center}
{
    This draft: \monthyeardate\today
}
\end{center}
\thispagestyle{empty}

\begin{abstract}
    \begin{spacing}{1.0}
    Analyzing a comprehensive news dataset, we document that joint news coverage triggers attention contagion, causing temporarily inflated valuations for affected stocks. Tracing SEC EDGAR visits from unique IPs, we provide direct evidence of attention spillovers between stocks. Stocks with greater joint news coverage exhibit increased Google search activity, higher contemporaneous returns, and subsequent reversals. Notably, aggregated joint news coverage strongly and negatively predicts future market returns. This relationship holds out-of-sample, persists after controlling for existing predictors and fundamental linkages, and intensifies during periods of heightened uncertainty or significant market frictions. Our findings indicate that attention contagion contributes to marketwide overvaluations.
    \medskip

     \medskip

    \end{spacing}
\end{abstract}

\end{titlingpage}

\setcounter{page}{1}
\renewcommand*{\thefootnote}{\arabic{footnote}}
\interfootnotelinepenalty=10000 %footnote in one page

\section{Introduction} \label{SecIntro}

An extensive body of literature suggests that investor attention plays a crucial role in shaping the trading activity and pricing dynamics of individual stocks. In particular, \cite{BarberOdean2008}, \cite{DaEngelbergGao2011}, and \cite{Barber2021} document that increased attention to individual stocks elevates their prices in the short term but predicts negative future returns.\footnote{Another line of research focuses on limited attention and shows that such frictions contribute to underreaction-related anomalies (\citealt{Hirshleifer2003, Hirshleifer2004, Peng2005, Cohen2008, dellavigna/pollet:09, HirshleiferLimTeoh2009, LiuPengTang2023, hirshleifer2025news}) as well as excessive return comovements (\citealt{PengXiong2006, Huang2019}).} However, while strong evidence exists at the individual stock level, cross-sectional predictors generally do not contain systematic information and are not reliable for time-series predictions (\citealt{Engelberg2023}). As a result, if the observed attention effects are predominantly stock-specific, their broader market implications may be limited because these effects can be diversified away. This raises a critical question: Does investor attention have meaningful implications for the overall market and for well-diversified investors?

This paper addresses this question by providing novel evidence of attention contagion in a cross-firm setting, demonstrating that investor attention has significant aggregate implications and contributes to marketwide overvaluation and subsequent corrections. Specifically, we examine the spillover of attention across firms through the lens of financial news covering multiple firms, a phenomenon we refer to as \textit{joint news coverage}. 
 
Combining data on financial news with unique IP access records of SEC EDGAR filings, we demonstrate that joint news coverage is associated with substantial attention spillovers across stocks. Specifically, we find that investors who follow one firm tend to expand their attention to other firms mentioned in the same news articles. Moreover, higher joint news coverage is associated with elevated contemporaneous returns for individual stocks, followed by significant reversals in the subsequent month. Importantly, the stock-level effects of joint coverage aggregate to the industry and market levels: our measure of aggregate joint news coverage negatively predicts future industry- and marketwide returns. The results are robust to a comprehensive set of alternative return predictors and are distinct from slow information diffusion driven by fundamental linkages across firms. These findings provide evidence consistent with the idea that media coverage generates attention contagion, triggers stock price inflation, and contributes to marketwide overvaluation.

This news-based approach is motivated by prior research showing that news coverage is a key driver of investor attention and is closely associated with individual stock mispricing.\footnote{See, for example,  \cite{Huberman2001}, \cite{BarberOdean2008}, \cite{Tetlock2011}, \cite{Hillert2014}, for these arguments. Additionally, media coverage can reduce information frictions by disseminating information to market participants \citep{FangPeress2009, Peress2014}.  Media reporting has also been shown to drive the trading of local stocks \citep{EngelbergParsons2011} and capital flows into mutual funds \citep{KanielParham2017}.} 
\label{Lit_Discussion} To examine investor attention in a multi-firm setting, our conceptual framework integrates insights from \citet{Merton1987}, \citet{BarberOdean2008}, \citet{DaEngelbergGao2011}, and \citet{Barber2021}. We rely on \citet{Merton1987} to understand attention spillovers and on the latter studies to draw pricing implications. As noted by \citet{Merton1987}, investors tend to overlook news about stocks they are unaware of or unfamiliar with, and as a result, avoid investing in such stocks. This behavior results in a firm's investor base comprising only a subset of potential market participants.

We therefore posit that joint news coverage of multiple firms directs investor attention to previously overlooked firms, creating attention spillovers across the covered firms and increasing the attention their stocks receive. Subsequently, as suggested by \citet{BarberOdean2008}, \citet{DaEngelbergGao2011}, and \citet{Barber2021}, attention-driven aggressive buying of these stocks leads to overvaluation, followed by subsequent reversals.

\begin{figure}[htp!]
    \centering
    \includegraphics[width = 0.9\linewidth]{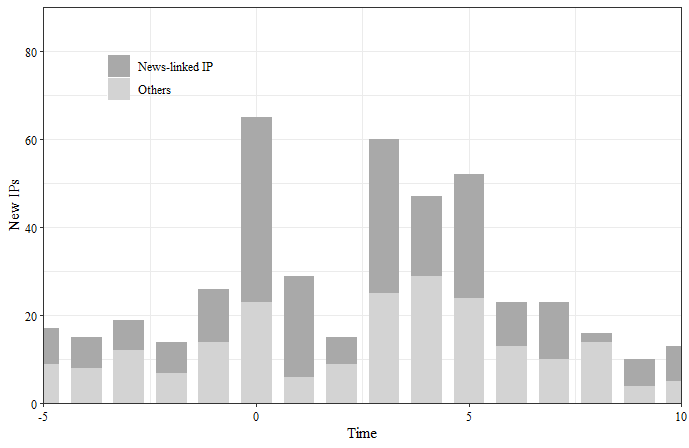} \caption{\textbf{EDGAR IP activities for NSC.} The figure plots the daily number of unique new IPs that accessed NSC's EDGAR filings over the [$-5, 10$] day window around April 22, 2016. We proxy for the (pre-event) investor base of a stock using the number of unique IP addresses that visited the stock's SEC filings in EDGAR over the past 18 months. A ``new" IP visit is defined as a unique IP address that accessed NSC's filings during the event window but had not accessed them in the preceding 18 months. These new IP addresses are categorized into two groups: News-linked IPs, which previously accessed filings of firms jointly mentioned in the same news article, and Others, which include all remaining new IPs.}
\label{Fig-Example}
\end{figure}

\label{Example_Discussion} Appendix \ref{App:Example} provides an illustrative example of attention spillover associated with joint news coverage.  On April 22, 2016, a  Reuters news headline featured four stocks: Microsoft (MSFT), Alphabet (GOOGL), Norfolk Southern Corp. (NSC), and BJ's Restaurants (BJRI). To track patterns in investor attention to NSC, a NYSE-listed railroad operator with a market capitalization of \$24.6 billion at the end of March 2016, we analyze unique IP activities accessing regulatory filings through the Securities and Exchange Commission (SEC)'s EDGAR (Electronic Data Gathering, Analysis, and Retrieval) system.\footnote{See \cite{LeeMaWang2015} and \cite{Drake2020} for studies using internet traffic data from the SEC's EDGAR website to infer information acquisition activities.} Figure \ref{Fig-Example} documents a sharp 261\% increase in new IP visits to the NSC filings --- defined as IP addresses that had not accessed the filings in the previous 18 months --- from a daily average of 18 during the 5-day pre-news window to 65 on the news release day. Notably, a significant proportion of these new IP visits are labeled as News-linked IP --- 42 out of the 65 --- originated from IP addresses that had previously accessed filings of firms jointly mentioned in the same news article, with 23 attributed specifically to prior visits to MSFT's filings.\footnote{Of the 353 new IPs that accessed NSC filings within a 10-day window spanning the event day and the subsequent nine days, over 50\% (184) originated from IPs that had previously accessed filings of firms jointly mentioned in the news. Among these, IPs previously accessing MSFT filings contributed approximately a quarter (86). A complete list of new IPs from MSFT during the post-event window is provided in Appendix Table \ref{AppTab-IPs}.}

This example suggests that joint news coverage of both NSC and MSFT likely directed the attention of MSFT investors, who were previously inattentive to NSC, toward NSC. Moreover, Figure \ref{Fig-App} shows that the cumulative return of NSC increases by more than 5\% on the event day, followed by an almost complete reversal over the subsequent 20 days. The pattern of return reversal following joint news coverage suggests the possibility that joint news coverage is associated with the overvaluation of affected stocks. \label{Example_Discussion_End}

$$[\text{Insert Figure \ref{Fig-Example} here.}]$$

In our empirical analysis, we formally examine the extent to which joint news coverage generates investor attention spillovers and influences stock returns. We begin with a comprehensive dataset comprising over 6.6 million news articles covering firms in the S\&P 500 index from January 1996 through December 2019. From this dataset, we construct a monthly cross-firm matrix of abnormal news coverage. In this matrix, the off-diagonal elements represent joint coverage, capturing the frequency with which a pair of firms is mentioned together in the same news article. The diagonal elements, by contrast, reflect self news coverage, where only one firm is mentioned in a given article.  

Building on the work of \citet{Banerjee2013, Banerjee2019}, who show that central nodes in an information network are more effective in spreading information, we hypothesize that attention spillovers are stronger from peer firms with high-centrality positions in the news coverage matrix to the focal firm. Accordingly, we define the extent to which a focal firm $i$ is subject to joint news coverage with its peers, JointNews$_i$, as the sum of abnormal joint news coverage across the peers, weighted by the centrality of each peer firm.

\label{Response_R7_1} We begin with the cross-sectional analysis of joint news coverage and investor attention. We first show that higher values of JointNews for a firm are associated with a significant increase in abnormal Google search volume for the stock, a widely used proxy for retail investor attention \citep{DaEngelbergGao2011}.

To further investigate attention spillovers at the individual-investor level, we analyze data on requests for firms' EDGAR filings from 5.97 million unique Internet Protocol (IP) addresses for the period January 2003 through June 2017. Within this dataset, we define a firm's investor base as the set of IPs that have previously downloaded the firm's filings. We then measure the number of unique new IPs accessing a focal firm's filings that overlap with the existing investor base of other firm(s) mentioned in the news (denoted as New IP$_c$). 

We find that increases in joint news coverage are associated with significant growth in new IPs. Specifically, a one standard deviation increase in a firm's joint news coverage is associated with a 15\% increase in visits to its EDGAR filings by new IPs that previously accessed the stock's news-linked peers. Notably, this effect is substantially larger than that of self-news coverage. These findings provide direct evidence consistent with the hypothesis that joint news coverage expands the investor bases of covered firms by driving attention spillovers.

We then examine the relationship between joint news coverage and stock returns, finding that a firm's joint news coverage is positively associated with returns in the current week but negatively associated with returns in the subsequent week and month. This dynamic is consistent with the hypothesis that joint news coverage triggers temporary overvaluation, followed by subsequent price corrections.

To explore the economic mechanism underlying our findings, we assess whether the return predictability of joint news coverage is driven by economic linkages among peer firms featured in the news or specific news topics discussed. We find that the results remain robust after controlling for various types of fundamental similarities between firms such as shared industry, production network,  technology, and analysts (see, e.g., \citealt{Moskowitz1999}, \citealt{Cohen2008}, \citealt{Menzly2010}, \citealt{Lee2019}, \citealt{Rapach2019}, and \citealt{Ali2020}). Furthermore, using the methodology proposed by \citet{Oster2019}, we compare coefficient estimates with and without controls. The proportional selection parameter and the absolute ``delta statistic" indicate that omitted variable bias is unlikely to account for our findings.

To further address the alternative explanation that our findings are driven by the content of news articles capturing important economic linkages between firms, we decompose our key variable, \text{JointNews}, into 15 topic-specific components, with topic definitions following \citet{Scherbina2015}, along with a residual orthogonal to these topic-specific measures. Our analysis indicates that the predictive power of the residual component of \text{JointNews} remains comparable to that of the original \text{JointNews} measure, suggesting that fundamental news concerning the firms is unlikely to be a major channel driving our results.

These results therefore suggest that our findings differ fundamentally from the prior literature, which documents \textit{positive} return predictability between fundamentally connected peers, consistent with underreaction to peer information. In contrast, we find a strong \textit{negative} relationship between the intensity of joint news coverage and future returns. Together with the user-level results, our findings are consistent with the hypothesis that joint news coverage triggers investor attention spillovers from connected firms to the focal firm due to non-fundamental reasons. These spillovers lead to higher valuations for the affected stock in the short term, followed by subsequently lower future stock returns.

We then examine whether the news-driven attention spillover effect, when aggregated, can influence aggregate returns. To this end, we aggregate the firm-level joint news measure into value-weighted industry-level and market-level measures, termed \text{JointNews$^{I}$} and \text{JointNews$^{M}$}, respectively. These measures capture the extent of joint news-driven attention spillovers within an industry or across the entire market. Similarly, we define \text{SelfNews$^{I}$} and \text{SelfNews$^{M}$} as the value-weighted averages of the firm-level self news coverage measure at the industry and market levels, respectively. Since self news is more likely to attract stock-specific attention rather than attention spillovers, whereas joint news is expected to generate attention spillovers across multiple stocks, we hypothesize that \text{JointNews$^M$} will exhibit a stronger association with aggregate returns compared to \text{SelfNews$^M$}.

To test this hypothesis, we compare the predictive ability of \text{JointNews$^I$} and \text{SelfNews$^{I}$} for industry returns, as well as \text{JointNews$^M$} and \text{SelfNews$^{M}$} for market returns, evaluate their relative effectiveness in generating investor attention and eliciting market price responses. 

We find that \text{JointNews$^I$} strongly negative predicts the one-month-ahead industry returns, and \text{JointNews$^M$} strongly and negatively predicts one-month-ahead market returns, with a sizable in-sample $R^2$ of 1.81\% and a substantial out-of-sample $R^2$ of 2.90\%. For a mean-variance investor, utilizing the information in \text{JointNews$^M$} leads to annualized certainty equivalent return (CER) gains of 4.23\% for a reasonable risk aversion. Moreover, portfolios formed based on \text{JointNews$^M$} achieve an annualized Sharpe ratio of 0.62, which is 15\% higher than the market Sharpe ratio of 0.53 for a buy-and-hold strategy.

\label{Response_R7_2} The predictability is robust to a large list of alternative predictors of market returns or variables that capture macroeconomic conditions. Specifically, we control for the 14 economic predictors identified by \citet{Goyal2008}, investor sentiment measures from \citet{Baker2006, Baker2007} and \citet{Huang2015}, and investor attention measures, including the FEARS index of \citet{DaEngelbergGao2015}, as well as measures from \cite{Chen2020}, \citet{LiuPengTang2023}, and \citet{Yuan2015}. In addition, we incorporate news-content-based predictors of market returns and measures of macroeconomic conditions, including the news tone measure from Thomson Reuters News Analytics and the topic-based measures proposed by \citet{Calomiris2019} and \citet{Bybee2021}. Finally, we control for signals of marketwide stock mispricing, including merger waves and IPO waves, as suggested by \cite{Shleifer2003} and \cite{Baker2002}. 
Joint news coverage exhibits strong predictive power both in- and out-of-sample and is robust across business cycles. Moreover,  the predictability persists  for at least three months, with a one-standard-deviation increase in \text{JointNews$^M$} reducing the aggregate stock market returns over the next six months by 0.57\% per month. 

This suggests that \text{JointNews$^M$} contains important information about future aggregate market returns that is not captured by the existing predictors. In contrast, \text{SelfNews$^{M}$} lacks significant predictive power for market returns. These results align with our hypothesis that joint news triggers a contagion of investor attention and therefore marketwide valuation increases, whereas self-news' impacts tend to be idiosyncratic. Our findings suggest that the stock-level effects of investor attention documented by \citet{BarberOdean2008}, \citet{DaEngelbergGao2011}, and \citet{Barber2021} extend beyond individual stocks --- periods of intensive joint news coverage can generate attention contagion, resulting in marketwide overvaluation and subsequent reversals.  

\label{Merton_Discussion} We also investigate whether our findings could be explained by a rational explanation, wherein increased investor attention improves risk-sharing and lowers a stock's required rate of return (\citealt{Merton1987}). To address this possibility, we conduct two additional analyses. The first analysis focuses on individual stocks and employs a direct measure of mispricing, as developed by \cite{Stambaugh2015}. The results show that stocks subject to greater joint news coverage are significantly more likely to be overvalued, providing evidence inconsistent with a purely rational, risk-based explanation.

The second analysis examines the relationship between aggregate joint news coverage and market returns in a time-series framework, focusing on the role of market frictions. According to the mispricing-based explanation, return predictability should be stronger during periods when arbitrage is costly and weaker when arbitrage costs are low. In contrast, the rational risk-based explanation suggests that predictability should persist regardless of arbitrage costs, as it reflects compensation for risk rather than temporary mispricing. Our findings reveal that \text{JointNews$_{M}$} significantly predicts negative market returns only during periods of high arbitrage costs, while the predictability becomes insignificant when arbitrage costs are low.

Together, this evidence from both cross-sectional and time-series analyses aligns more closely with the mispricing mechanism rather than the improved risk-sharing mechanism. These results suggest that the return predictability associated with joint news coverage is driven by temporary overvaluation and subsequent reversals, rather than by a rational adjustment for risk.\label{Merton_Discussion_End}

In additional analysis, we follow \cite{Peress2020} and use episodes of sensational news \citep{Eisensee2007} that are unrelated to the financial market as exogenous shocks to \text{JointNews$^M$}. We first establish that, all else being equal, sensational news significantly reduces \text{JointNews$^M$}, confirming that these episodes distract media attention and lower joint news coverage. We then use sensational news as an instrumental variable for \text{JointNews$^M$} to predict future market returns. We find that the relationship between \text{JointNews$^M$} and market returns remains strongly negative, with a one standard deviation increase in the predicted \text{JointNews$^M$} significantly lowering the following month's market return by 43 basis points. Thus, the two-stage instrumental variable analysis suggests that random exogenous fluctuations in \text{JointNews$^M$} account for a substantial portion of the variable's predictive power, beyond what fundamental linkages alone could explain. 

Our findings directly contribute to the literature on investor attention, which demonstrates that investor attention has a significant influence on the prices of individual stocks.\footnote{For example, \citet{DaEngelbergGao2011} show that retail investor attention positively predicts short-term stock returns and a subsequent reversal, while \citet{BenRephael2017} and \citet{BenRephael2021} find institutional attention facilitates a permanent and efficient reaction to information. \cite{Fedyk2021} shows that prominent news on the Bloomberg terminal attracts more trading and price responses. \citet{Bali2019} and \citet{Atilgan2020} show that attention enhances investors' attraction to lottery stocks. \citet{ChenLiYu2020} find lead-lag effects in returns for stocks displayed together. More recently, \cite{Cookson2023} and \cite{Barber2021} show that fintech brokerages and social interactions influence investor attention and contribute to stock returns. Other papers explore the implications on return comovements \citep{PengXiong2006, Drake2016, Huang2019}.} What is novel in our paper is that we highlight that joint news coverage can make those individual effects contagious and generate aggregate impacts on the overall market. Two recent papers also examine the relationship between investor attention and market returns. \citet{Chen2020} find that a common component of 12 investor attention proxies has power in predicting market returns, while \citet{Da2024} find that retail and institutional attention have distinctively different predictive powers.  By identifying attention spillovers through joint news coverage, this paper uncovers a critical channel through which investor attention influences aggregate market outcomes. \label{Response_R7_3} 

Our paper also relates to recent studies that use common news coverage to identify fundamental linkages among companies. For example, \cite{LeeMaWang2015} use EDGAR co-search activities to identify fundamental similarities between firms. \cite{Scherbina2015} find that returns of peer firms that share common news coverage can help predict focal firm's stock returns and associate the predictability with slow information diffusion. \cite{Schwenkler2020} use news-implied networks to identify links between distressed firms and find predictable stock returns and credit downgrades. \cite{Li2022} use cross-firm networks generated from news to estimate a spatial factor model to capture equity return comovements. As we explained earlier, our findings differ substantially from prior studies that focus on fundamental linkages between firms, and our evidence highlights the role of attention spillover due to non-fundamental reasons. Another important difference between our findings and the aforementioned studies is that, unlike our variable, there is no evidence that these alternative, fundamental-related patterns are relevant for aggregate returns. These key differences underscore the distinct mechanism underlying our results, showing that common news coverage generates attention contagion and return predictability, both at the individual stock level and marketwide, in a way that cannot be explained by fundamental linkages.

More broadly, our paper contributes to the growing body of literature that demonstrates the significant role media and news play in financial markets. Several papers study the relationship between news and market returns.  For example, \citet{Tetlock2007} and \citet{Garcia2013} find that media pessimism predicts aggregate stock market returns. \citet{Froot2017} document a ``reinforcement effect'' between returns and media-measured sentiment. \citet{Glasserman2019} show that the ``unusualness" of news with negative sentiment forecasts market stress, while \citet{Calomiris2019} find that news-based word flow measures predict risk and returns. \citet{Bybee2021} show that news attention closely tracks a wide range of economic activities and explains 25\% of aggregate stock market returns. 

Our results remain robust after accounting for the key variables used in these studies. More importantly, we provide new insight into one of the channels through which news affects investor behavior and prices. Our use of user-level IP activities provides a microfoundation of the observed attention-return patterns and bridges the gap between firm-level and aggregate effects of news on stock markets. Our findings also open up new avenues for future research. In particular, survey evidence suggests that investors' beliefs play a central role in financial markets and the beliefs are characterized by large and persistent heterogeneity (\citealt{Giglio2021a, Giglio2021b}). Exploring how investor attention shapes beliefs and the role of news media or social media in directing that attention could be a fruitful direction to pursue.

The paper is organized as follows. Section \ref{SecHypo} explains the conceptual framework and develops the attention spillover hypotheses. Section \ref{SecData} describes the data and the construction of variables. Section \ref{SecCX} presents the novel direct evidence of attention spillover and examines firm-level return predictability. Section \ref{SecTS} extends the analysis to the market level, examining market-level return predictability. Section \ref{SecFurther} investigates whether our findings could be explained by a rational explanation and whether our results may be driven by omitted variables. Section \ref{SecConcl} concludes.

\section{Hypothesis Development} \label{SecHypo}
		
In this section, we establish the conceptual framework and develop the attention spillover hypothesis in a cross-firm setting.

As discussed in the introduction, prior research demonstrates that media coverage of firms elicits significant investor reactions. However, not all news garners equal attention from all investors. The concept of limited investor recognition, introduced by \citet{Merton1987}, posits that investors are only aware of or familiar with a restricted set of stocks. As a result, investors may overlook news related to unfamiliar stocks and refrain from investing in them. Building on this premise, we hypothesize that news articles mentioning multiple stocks generate a cross-firm attention spillover. Specifically, investors following one stock, after encountering such articles, begin to pay attention to other mentioned stocks, potentially becoming investors in these additional securities. This attention spillover broadens the investor base of all covered stocks, thereby increasing overall investor attention to those stocks.

\begin{figure}[htp!]
  \centering
  \includegraphics[width=0.7\textwidth]{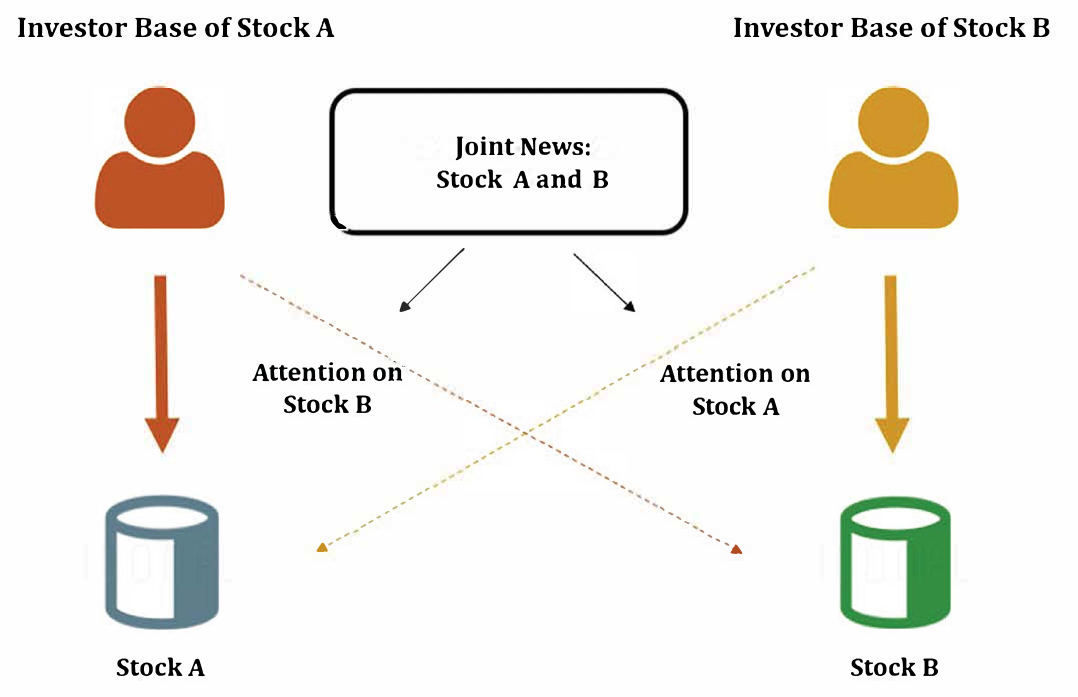}
\end{figure}

We illustrate the intuition with a simple example. Consider two stocks, A and B, and focus on the corresponding non-overlapping sets of their investor bases,  $I_A$ and $I_B$.\footnote{The spillover effect applies only to the non-overlapping investor base, so we omit the overlapping set in our discussion for brevity.} That is, investors in $I_A$ are only aware of stock A and are only attentive to news about A, and investors in $I_B$ are only aware of stock B and are only attentive to news about B. While news articles that cover only one stock, A, will only be read by $I_A$ investors, articles that mention both stocks (\textit{joint news}) attract attention from both $I_A$ and $I_B$ investors. Therefore, the arrival of joint news and $I_A$ investors' consumption of the news make $I_A$ investors aware of stock B and subsequently cause them to start paying attention to B. We refer to this effect as a spillover of investor attention from stock A to stock B. As a result, both $I_A$ and $I_B$ investors are attentive to stock B. Similarly, investor attention spills over from B to A, and stock A now receives the attention from both $I_A$ and $I_B$ investors. 

Regarding the implications for individual stock returns, motivated by \citet{BarberOdean2008, DaEngelbergGao2011, Barber2021}, we postulate that greater attention triggers more attention-based buying, resulting in temporarily high valuations and subsequent low returns for affected stocks. We summarize hypothesis \ref{hyp1} as follows:

\begin{hyp} \label{hyp1}
The arrival of a news article covering multiple stocks generates a spillover of investor attention across the stocks mentioned in the article. Stocks that receive high volumes of joint news coverage tend to experience high valuations but subsequently yield low returns.
\end{hyp}

More importantly, during periods when numerous joint news articles are released, the attention spillover effect becomes widespread, generating systematically high valuations in market prices, followed by lower future market returns. The intensity of cross-firm attention spillover depends on a firm's importance in the cross-firm network of news coverage. Building on the influential work of \citet{Banerjee2013, Banerjee2019}, who find that central nodes of a network are more influential in information transmission, we assign higher weight to stocks corresponding to high-centrality nodes in the cross-firm network of news coverage. That is, the attention spillover from a connected firm to a focal firm is stronger if the connected firm occupies a more central position in the network. Hence, we state the following hypothesis:

\begin{hyp} \label{hyp2}
During periods when a large number of joint news articles are released, the cross-firm spillover of investor attention aggregates and generates high market prices, resulting in low future market returns. This effect is stronger when the spillover originates from higher-centrality nodes of the cross-firm network of news coverage.
\end{hyp}

Regarding self-news, due to its idiosyncratic nature and the absence of an attention spillover effect, we expect self-news to have a weak association with market returns.

\section{Data and Variables} \label{SecData}

In this section, we describe how we construct the joint news coverage measures and describe our news data, followed by descriptions of the stock return data and other predictors.

\subsection{Cross-Firm news coverage matrix}  \label{subsec:News_matrix}

We measure news coverage across firms by analyzing an extensive database of 6,607,068 U.S. news stories covering firms in the S\&P 500 index from January 1996 through December 2019.\footnote{The number of news articles is comparable to that in Table 1 of \citet{ChenKellyXiu2022}, where they use the same Thomson Reuters News Archive dataset with the period ranging from January 1996 to June 2019.} Using Thomson Reuters's News Analytics and Thomson Reuters news archive, we extract news date, story identifier, the firms mentioned in the news, and the full text of the news. We classify news articles into two categories: 1) joint news, which mentions at least two firms, and 2) self news, which mentions only one firm. Following these criteria, 1,367,755 stories are classified as joint news, of which 695,804 stories mention exactly two firms and 671,951 stories mention more than two firms. 

We construct a monthly cross-firm news coverage matrix $\mathbf{W}_t$ for each month $t$:
\begin{equation} \label{eqNewsCov}
\mathbf{W}_t = \kbordermatrix{
	& stock_1 & stock_2 & \cdots & stock_{N}\\
	stock_1 & w_{11,t} & w_{12,t} & \cdots & w_{1N,t} \\
	stock_2 & w_{21,t} & w_{22,t} & \cdots & w_{2N,t} \\
	\vdots  & \vdots & \vdots & \ddots & \vdots \\
	stock_N & w_{N1,t} & w_{N2,t} & \cdots & w_{NN,t}},
\end{equation}
where $N$ is the total number of firms in our sample; the off-diagonal element, $w_{ij,t}$ with $i \neq j$,  corresponds to the number of news articles that mention both firms $i$ and $j$ (\textit{joint coverage}); and the diagonal element, $w_{ii,t}$, is the number of news articles that mention only firm $i$ (\textit{self coverage}).

As hypothesized, more-central firms in the cross-firm network of news coverage tend to have a larger investor base and thus are more likely to generate a greater attention spillover effect onto other firms through joint news coverage. We measure firm $i$'s centrality in the news coverage matrix using the eigenvector centrality of node $i$.\footnote{The eigenvector centrality of firm $j$ is the $j$th element of the principal right eigenvector of the adjacency matrix and is proportional to the sum of the centrality scores of its direct neighbors and therefore accounts for the transmission of signals along longer paths and walks. As suggested by \citet{Newman2010}, eigenvector centrality describes the informativeness of the links in an information network.} We further detrend the news coverage series and define $\widetilde{w}_{ij,t}$ as the logarithm difference between $w_{ij,t}$ and its past six-month median.

The abnormal joint news coverage for a given firm $i$, JointNews$_i$, is the sum of abnormal joint news coverage across firm pairs $(i,j)$, $j = 1, \cdots, J$, weighted by the centrality of $j$. 

We aggregate the firm-level measure to the market level and define JointNews$^M$ as the market capitalization-weighted average of JointNews$_i$. For robustness, we construct an alternative measure, JointNews$^M_{ew}$, as the equal-weighted average of joint news coverage. We similarly construct an aggregated self-news coverage measure, SelfNews$^M$, by value-weighting abnormal self-news coverage across firms and compare the measure's effects with that of JointNews$^M$.

\subsection{Stock returns and other predictors} \label{subsec:Data_other}

The firm-level stock data are sourced from CRSP, while accounting and financial statement variables are obtained from the merged CRSP-Compustat database. We consider the incremental contribution of the joint news relative to existing predictors of firm-level and market returns. Specifically, for predicting firm-level returns, we follow \citet{DaEngelbergGao2011} and \citet{BenRephael2017} to include the following variables: the logarithm of market capitalization (ln(Size)), the logarithm of book-to-market ratio (ln(BM)), idiosyncratic volatility (IVOL), standardized abnormal turnover (AbnTnvr) as in \cite{Chordia2007}, advertisement expenses to sales ratio (AdExp/Sales), and log number of analyst coverage (ln(1+Analyst)). 

In terms of predicting market returns, we follow \citet{Goyal2008} and include the following 14 economic predictors: log dividend-price ratio (DP), log dividend yield (DY), log earnings-price ratio (EP), log dividend-payout ratio (DE), stock return variance (SVAR), book-to-market ratio (BM), net equity expansion (NTIS), Treasury bill rate (TBL), long-term bond yield (LTY), long-term bond return (LTR), term spread (TMS), default yield spread (DFY), default return spread (DFR) and inflation rate (INFL).

Next, we obtain the negative news tone (News Tone) measure from the Thomson Reuters News Analytics. We also include the investor sentiment measures proposed by \citet{Baker2006, Baker2007} (Sent$^{BW}$) and \citet{Huang2015} (Sent$^{PLS}$). Additionally, we collect several alternative investor attention measures, including the Financial and Economic Attitudes Revealed by Search (FEARS) index by \citet{DaEngelbergGao2015} and the composite investor attention measure (Attn$^{PLS}$) by \cite{Chen2020}, and construct the abnormal Google search volume for major stock indices (ASVI) following \citet{LiuPengTang2023} and the nearness to the Dow historical high measure (RcrdHigh) which transforms the daily indicator in \citet{Yuan2015} into a monthly continuous variable following \citet{Chen2020}.\footnote{We obtained the economic predictor data from Amit Goyal's website, the Baker-Wurgler investor sentiment index from Jeffery Wurgler's website, and the PLS sentiment index from Dashan Huang's website. The FEARS index was retrieved from Zhi Da's website, and the PLS investor attention index from Guofu Zhou's website. We are grateful to the authors for generously making these datasets publicly available.} Furthermore, we collect the number of Initial Public Offerings (IPO) from Jay Ritter's website and the number and gross value of Mergers and Acquisitions (M\&A) events (MA$_n$ and MA$_v$) from the IMAA institute. All the definitions of the variables used in the paper are summarized in Table \ref{Tab-VarDef}. 

\subsection{Summary statistics} \label{subsec:SumStats}

Table \ref{Tab-SummaryStats} reports the summary statistics for the firm- and market-level variables used in the paper in Panel A and Panel B, respectively. Notably, JointNews$^M$ has an AR(1) coefficient of 0.36 and is much less persistent than most of the other predictors. The monthly market return has a mean of 0.7\%, a standard deviation of 4.3\%, and an AR(1) coefficient of 0.04. These summary statistics are consistent with the literature. Table \ref{Tab-Corr} shows that the correlation coefficients between JointNews$^M$ and other predictors are not high, suggesting that JointNews$^M$ differs from the other predictors and may contain information for predicting market returns that is incremental to the predictors established in prior literature.   

$$[\text{Insert Table \ref{Tab-SummaryStats} here.}]$$

\begin{figure}[htp!]
	\begin{center}
		\includegraphics[width = 0.95\textwidth]{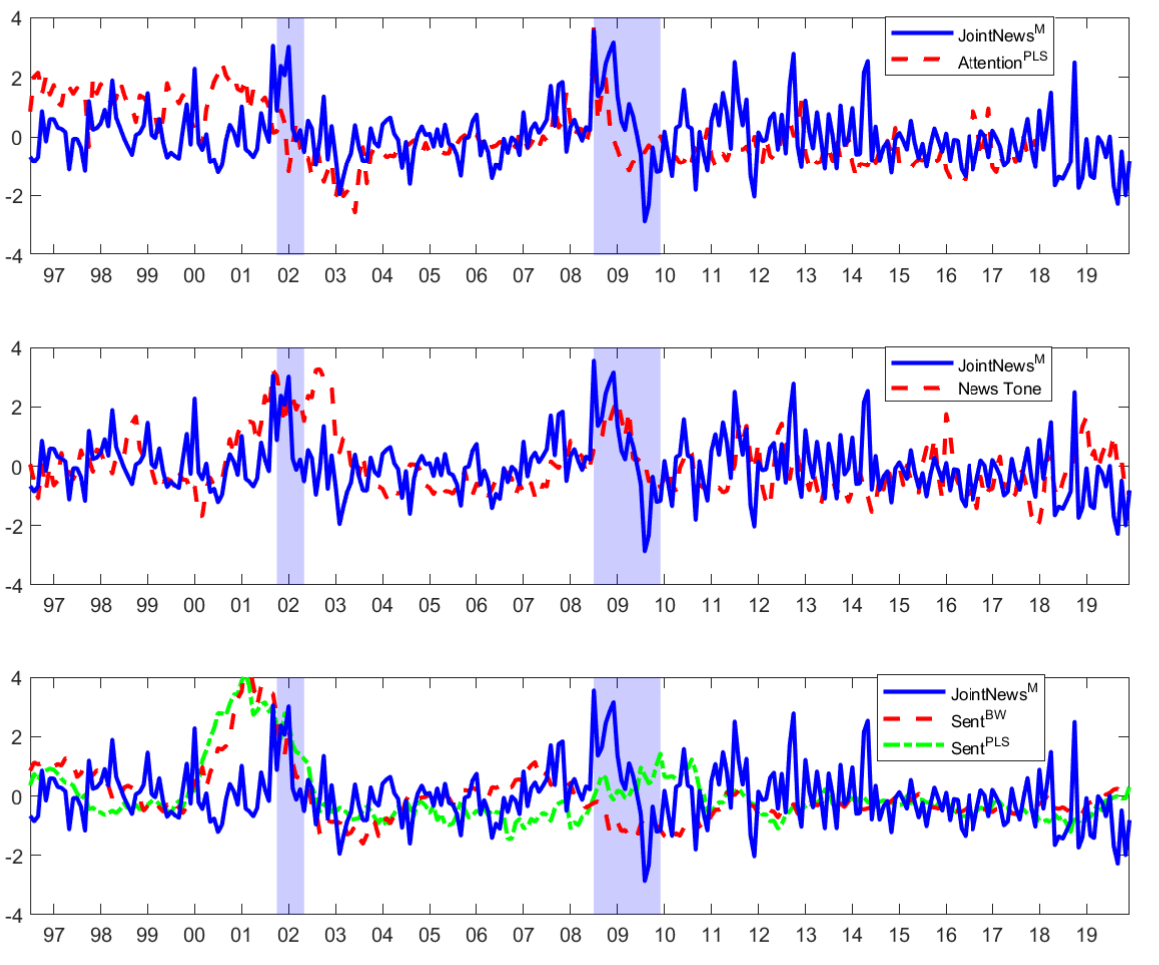}
	\end{center}
	\caption{\textbf{Joint News.}The figure plots the standardized time series of \text{JointNews$^M$} with composite attention index, News Tone, and investor sentiment indices. Panel A plots \text{JointNews$^M$} (solid line) and composite attention index by PLS \citep{Chen2020} (dashed line). Panel B plots \text{JointNews$^M$} (solid line) and negative news tone measure from Thomson Reuters News Analytics (dashed line). Panel C plots \text{JointNews$^M$} and investor sentiment measures \citet{Baker2006} (Sent$^{BW}$, dashed line) and \citet{Huang2015} (Sent$^{PLS}$, dot-dashed line). All indices are standardized to have a mean of zero and a standard deviation of one. The shaded periods correspond to NBER-dated recessions. The sample period spans from July 1996 to December 2019.} \label{Fig-JointNews}
\end{figure}

Given that our joint news measure is related to news, investor attention, and sentiment, we compare the standardized time series of these measures in Figure \ref{Fig-JointNews}. The figure shows that JointNews$^M$ (blue solid line) is distinctly different from the composite investor attention Attention$^{PLS}$ of \cite{Chen2020}, the News Tone of \citet{Tetlock2008}, and the market sentiment measures, Sent$^{BW}$ and Sent$^{PLS}$, of \citet{Baker2006} and \citet{Huang2015}. There are also considerable temporal variations in JointNews$^M$ compared with these other return predictors. 

$$[\text{Insert Figure \ref{Fig-JointNews} here.}]$$

\section{Cross-sectional Analysis} \label{SecCX}

In this section, we first provide direct evidence that joint news coverage of firms triggers cross-firm investor attention spillover. Specifically, we examine the relationship between joint news coverage and Google search activities, as well as EDGAR search activities by unique IP addresses. Next, we investigate firm-level return predictability using joint news and explore whether the predictive power of joint news is driven by fundamental-related channels. Finally, we test industry-level return predictability by aggregating joint news from the firm level to the industry level.

\subsection{Joint news and attention spillovers} \label{subsec:XS_IP}

To investigate the relationship between joint news coverage of firms and investor attention to these firms, we first examine a commonly used proxy of firm-level investor attention, based on Google search activities. Following \citet{DaEngelbergGao2011}, we measure a stock's abnormal retail attention as the abnormal search volume (ASV) of the stock, which is the percentage change between Google's daily Search Volume Index (SVI) for a stock and its past one-year mean, excluding the most recent month.\footnote{The SVI is a relative search popularity score, defined on a scale of 0 to 100, based on the number of searches for a term relative to the total number of searches for a specific geographic area and a given period. We focus on searches made on weekdays in the U.S. market. We manually screen all tickers to select those that do not have a generic meaning (e.g., ``GPS" for GAP Inc., ``M" for Macy's) to ensure that the search results we obtain are truly for the stock and not for other generic items or firm products.} 
	
We analyze the relationship between ASV and the joint news and self news measures using the following panel regression:
\begin{align} 
    \text{ASV}_{it} = \beta_0 & + \beta_1 \cdot \text{JointNews}_{it} + \beta_2 \cdot \text{SelfNews}_{it} + \gamma \cdot X_{it} + \alpha_i + \lambda_t + \varepsilon_{it},  \label{eq:ASV}
\end{align}
where JointNews and \text{SelfNews} are the abnormal joint news and self news coverages, respectively. $I_{\text{SelfNews}}$ is the self news indicator that equals one if  \text{SelfNews} belongs to the top tercile of the cross-sectional distribution, and zero otherwise. $\alpha_i$ and $\lambda_t$ are firm and month fixed effects, respectively. $X$ is a set of control variables following \citet{DaEngelbergGao2011} and \citet{BenRephael2017}, including log firm size, log book-to-market ratio, idiosyncratic volatility, abnormal turnovers, advertisement expenses to sales ratio, and log analyst coverage. All independent variables are standardized to have a mean of zero and a standard deviation of one to facilitate comparison of their economic magnitudes.

$$[\text{Insert Table \ref{Tab-EDGARIP} here.}]$$

The results, reported in Table \ref{Tab-EDGARIP}, columns 1 and 2, show that increased joint news coverage significantly raises abnormal Google search activities for the firm. The coefficient of 0.790 for JointNews in column 2 indicates that a one standard deviation increase in joint news is associated with a 0.79 increase in ASV, representing a 79\% increase in search volumes relative to the previous 12-month mean. In contrast, a one-standard-deviation change in Selfnews in column 2 can only boost the abnormal Google search volume by 29\%. The signs of the coefficients of the control variables, $\ln(\text{Size})$, $\ln(1+\text{Analyst})$, $\text{AdExp}/\text{Sales}$, align with those in \citet{DaEngelbergGao2011}. Our findings suggest that joint news coverage attracts the attention of new investors to the focal firm's news articles that they would otherwise overlook. 

Next, we utilize granular, individual-investor-level data to gain a deeper understanding of attention dynamics in response to news. Motivated by studies that show EDGAR downloads are associated with information acquisition activities by investors \citep[see, for example,][]{LeeMaWang2015, ChenCohen2020}, we examine downloads of 10-K and 10-Q filings on the EDGAR website by unique IP addresses.

Since 1996, the Securities and Exchange Commission (SEC) has required all public domestic companies to submit their filings electronically via the EDGAR website. From the website, we obtain a sample consisting of over 230 million document requests from 5.97 million unique IP addresses for the period from 2003 to 2017.
We obtain the IP address of the EDGAR user, the filing firm, and the time stamp for each request. For each month, we proxy for a firm's investor base with the set of IPs that downloaded a firm's EDGAR filings in the past 18 months.\footnote{Users are partially anonymized as the EDGAR log files show the first three octets of the IP address and replace the fourth with a unique string. Following previous studies \citep[e.g.,][]{LeeMaWang2015, Drake2020, Ryans2017, Li2021}, we remove requests made by robots or automated webcrawlers by removing the records of users that download more than 50 unique firms' filings per day and keep only the successful request records (code = 200) and exclude records that refer to an index (idx = 1) since the index pages provide links to the firm's filings instead of the filings themselves. The choice of an 18-month window to define the existing investor base follows from the literature convention of allowing a sufficient window after the end of the fiscal year (up to 18 months) for the information in annual reports to become publicly available \citep[e.g.,][]{Fama1993, Hou2015}. The set of investors as captured by EDGAR IP activities is likely an important component of a firm's potential investors, although we acknowledge that the EDGAR-based set is only a subset of a firm's investor base. Hence, the results that we document likely capture a lower bound of the effect of joint news on investor attention.} 

To analyze joint news coverage and investors' information acquisition activities in EDGAR, we replace the dependent variable in Equation \ref{eq:ASV} with $\ln(\text{New IP}_{it})$, the logarithm of the number of unique new IP addresses accessing firm $i$'s EDGAR filings in month $t$ and estimate the following panel regression:
\begin{align} \label{eqNewIP}
    \ln(\text{New IP}_{it}) = \beta_0 & + \beta_1 \cdot \text{JointNews}_{it} + \beta_2 \cdot \text{SelfNews}_{it} + \gamma \cdot X_{it} + \alpha_i + \lambda_t + \varepsilon_{it},
\end{align}
where $X$ is the same set of control variables as in Equation (\ref{eq:ASV}), and we include firm and month fixed effects. 

Columns 3 and 4 of Table \ref{Tab-EDGARIP} show that both JointNews and SelfNews significantly increase EDGAR downloading activity by new IP addresses. The primary variable of interest, \text{JointNews}, has a coefficient representing the semi-elasticity of the number of new IPs with respect to joint news coverage. The coefficient of 0.034 on \text{JointNews} in column 4 indicates that a one-standard-deviation increase in \text{JointNews} is associated with a 3.4\% increase in the number of new IPs downloading a firm's EDGAR filings, holding other factors constant. While this effect is not large in magnitude, it is comparable to other firm characteristics such as analyst coverage or the book-to-market ratio.\footnote{\text{SelfNews} exhibits a similar effect, with a coefficient of 0.030. The result is consistent with \cite{Box2018}, which argues that firm-specific (self) news can trigger an attention spillover when it is similar to other news and is thus likely to be read together with other news. However, we note that such spillover effects are more limited compared to the effect of \text{JointNews}, as we later show that \text{SelfNews} is no longer significant in predicting industry-wide or marketwide returns, whereas \text{JointNews} remains highly significant.}

To further investigate how much of the new IP visits are attributable joint news coverage, we examine $\text{New IP}_{it}$, the number of unique new IP addresses accessing firm $i$'s EDGAR filings in month $t$ that had previously accessed the filings of other firms covered by the joint news.
This measure proxies for the extent to which joint news triggers a spillover of investor attention across the covered firms and quantifies how joint news coverage expands the investor base of the focal firm. 
The monthly averages of ln(New IP) and ln(New IP$_c$) are 5.587 and 3.942,  corresponding to non-logarithmic values of 267 and 51.5 per month, respectively.\footnote{For comparison, \citet{Drake2016} reports that the average daily number of EDGAR requests per firm is 20.} These figures indicate that a substantial fraction of new IPs, approximately 19.3\%, originate from joint news connections.

Table \ref{Tab-EDGARIP}, columns 5 and 6, present the result with the logarithm of New IP$_c$ as the dependent variable. The coefficient on JointNews is positive and highly significant. For economic magnitude: the estimate (0.150 in column 6) implies that a one-standard-deviation increase in JointNews is associated with a substantial 15\% increase in EDGAR downloads by new IPs originating from the investor bases of connected firms.\footnote{\text{SelfNews} also increases new IP activity, albeit with a much smaller coefficient of 0.034. A possible explanation is that exclusive news coverage of a firm may capture the attention of investors who have previously accessed the filings of firms mentioned in other news articles for the focal firm released during the same month.} 

Taken together, these findings are consistent with our first hypothesis that joint news coverage induces a substantial attention spillover across covered firms, significantly increasing overall investor attention to these firms. \label{subsec_end:XS_IP}

\subsection{Joint news and stock returns} \label{subsec:XS_Return}

Building on the finding that joint news coverage significantly increases investor attention to covered stocks, we next explore the asset pricing implications of this effect. Drawing on the insights of \citet{BarberOdean2008}, \citet{DaEngelbergGao2011}, and \citet{Barber2021}, our first hypothesis proposes that joint news coverage increases investor attention to covered stocks, resulting in temporarily high valuations that are subsequently followed by lower future returns for the affected stocks.

We test the hypothesis by estimating the following panel regressions:
\begin{align}
    R_{i, t\text{ or }i,[t+1:t+h]} = \beta_0 &+ \beta_1 \cdot \text{JointNews}_{it} + \beta_2 \cdot \text{SelfNews}_{it} + \gamma \cdot X_{it}  + \varepsilon_{i,t \text{ or }i,[t+1:t+h]}, \label{reg01}
\end{align}
where the dependent variable is the contemporaneous weekly stock returns, $R_{i,t}$  or future cumulative returns (in basis points) for up to four weeks, $R_{i,[t+1;t+h]}$, with $h$ varying from 1 to 4. JointNews and SelfNews are joint and self news coverage variables, and $X$ is the set of control variables defined in Equation (\ref{eq:ASV}). We also include firm and week fixed effects to control for unobserved heterogeneity across firms and over time. We exclude stocks with average prices below \$5 in the prior week to ensure that the results are not driven by small illiquid stocks. Following \citet{Thompson2011}, we compute the firm and week double-clustered standard errors to account for potential cross-firm and cross-time error correlations. 

$$[\text{Insert Table \ref{Tab-XSRet} here.}]$$

Results in Panel A, Column 1 of Table \ref{Tab-XSRet} document the contemporaneous relationship between weekly joint news coverage and stock returns. The significantly positive coefficient on \text{JointNews} suggests that stocks experiencing greater joint news coverage in a given week earn higher returns during that week. In contrast, Columns 2 and 3, which examine one-week-ahead returns and four-week cumulative returns, respectively, report significantly negative coefficients on \text{JointNews}.

In terms of economic magnitude, a one-standard-deviation increase in \text{JointNews} corresponds to a 2.3 basis point increase in same-week returns, followed by a 5.9 basis point decrease in cumulative stock returns over the subsequent month. These findings indicate that while joint news coverage is associated with temporarily elevated valuations, these effects reverse over time.

Given that most return predictors are reported at a monthly frequency, we follow the convention in the return predictability literature and conduct all subsequent analyses using monthly regressions, unless otherwise stated. Panel B of Table \ref{Tab-XSRet} presents results from monthly panel regressions of future returns on joint news coverage and the same set of control variables used in the weekly regressions, showing that firms with greater joint news coverage tend to experience lower future returns. For instance, the coefficient of $-10.29$ in Column 1 indicates that a one-standard-deviation increase in a firm's abnormal joint news coverage lowers its next-month return by approximately 10.3 basis points. Additionally, we find that \text{JointNews} exhibits significantly negative predictive power over short horizons, but its effect diminishes over longer horizons.

The observed pattern of a positive contemporaneous return relationship followed by subsequent reversals is consistent with our first hypothesis that joint news coverage generates temporary price pressure, resulting in high valuations that later revert. \label{subsec_end:XS_Return}

\subsection{Alternative explanations --- fundamental linkages} \label{subsec:XS_EconLink}

The evidence of a negative relationship between joint news coverage and future stock returns, combined with findings based on EDGAR IP accesses and Google searches, supports the hypothesis that joint news coverage triggers attention spillovers, which subsequently drive temporary stock overvaluation. Given that joint news may also highlight important economic linkages between jointly covered firms, we next investigate whether the return predictability of joint news coverage can be attributed to the predictability of economically connected peer firms or news contents that feature such linkages.

Prior research has examined information spillovers from peers with shared analyst coverage \citep{Ali2020}, industry affiliation \citep{Moskowitz1999}, geographic proximity \citep{parsons2020geographic}, customer-supplier relationships \citep{Cohen2008}, and technological similarity \citep{Lee2019}. These studies show that fundamentally linked firms exhibit greater similarities, and when investors fail to fully recognize these similarities, lead-lag relationships emerge between the returns of focal firms and their peers.

However, the return predictability associated with fundamentally linked peer effects differs conceptually from the joint coverage effects we document in several important ways. First, prior studies find that return predictability from fundamentally linked peers is directional --- lower peer returns predict lower focal firm returns, and vice versa. In contrast, our results show that joint news coverage intensity predicts future returns regardless of the sentiment of the news. Second, while fundamentally linked peer return predictability reflects investor underreaction and shows no evidence of return reversals, we find that joint news coverage is associated with contemporaneous high returns followed by subsequent return reversals. Third, as discussed in Section \ref{SecTS}, fundamentally linked peer return predictability lacks aggregate market implications, whereas aggregate joint news coverage significantly predicts future market returns.

To formally test whether our findings are attributable to fundamentally linked peer returns, we calculate connected-firm portfolio returns (CFRet$^{Fundamental}$), where \textit{Fundamental} refers to linkages based on shared analyst coverage (ana, \citealt{Ali2020}), customer-supplier relationships (cus, \citealt{Cohen2008}), industry affiliation (ind, \citealt{Moskowitz1999}), technological similarity (tech, \citealt{Lee2019}), and geographic proximity (geo, \citealt{parsons2020geographic}), following the original studies. We then augment Equation \ref{reg01} by including these peer returns as additional control variables.

$$[\text{Insert Table \ref{Tab-XSRet-Peer} here.}]$$

Table \ref{Tab-XSRet-Peer} presents monthly panel regressions of one-month-ahead returns on lagged joint news coverage. The JointNews coefficients remain significantly negative and similar in magnitude to those in Table \ref{Tab-XSRet}.\footnote{The weaker performance of these control variables in predicting future returns primarily reflects our sample, which is limited to S\&P 500 stocks, where underreaction to peer information is less pronounced. Appendix Table \ref{AppTab-CFRet}, which extends the analysis to all CRSP common stocks, replicates the findings documented in the prior literature.} These findings indicate that joint news coverage predicts returns through channels that are distinct from fundamentally-linked peer effects. 

We further rule out the possibility that our results of joint news coverage and future returns are due to the slow diffusion of fundamental information by exploring the contents of news. This analysis is motivated by \cite{Scherbina2015}, who perform content analysis of common news stories to identify information about economic linkages between firms and find return predictability of news-linked peer stocks.  

Following \cite{Scherbina2015}, we classify news articles into 15 topics: supply chain (Sply), partnerships (Ptnr), M\&A (MA), parent/subsidiary relations (Prnt), legal (Legal), regulation (Regl), labor, production and infrastructure (Lbr), executive compensation and corporate governance (CGvn), management news (Mgmt), common customer (Cstm), cross-investment (Ivst), natural resources (NRes), energy (Engy), technology (Tech), and geopolitical (Gptc). 

We then construct topic-specific joint news coverage for each firm. Specifically, for each news topic (e.g., supply chain), we compute the monthly cross-firm news coverage matrix as defined in Equation (\ref{eqNewsCov}) and calculate the abnormal joint news coverage for a given firm $i$ for a specific topic, denoted as \textit{JointNews}$_i^\textit{topic}$, which captures topic-specific information diffusion between linked firms. We then orthogonalize \textit{JointNews} with respect to the 15 topic-specific abnormal joint news coverage measures (\textit{JointNews}$_i^\textit{Topic}$) using the following regression:
\begin{equation} \label{eqResid}
    \text{JointNews}_{it} = \beta_0 + \sum_{s \in \{\textit{Topic}\}} \beta_s \cdot \text{JointNews}_{it}^{s} + \varepsilon_{it},
\end{equation}
and obtain the residuals $\widehat{\varepsilon}_{it}$. By construction, the residual from the above regression decomposition measures joint coverage that is not captured by the fundamental news contents. We then estimate a modified version of Equation \ref{reg01} by replacing JointNews$_i$ with either one of the topic-specific joint news coverage variables or the residual joint news coverage.

$$[\text{Insert Table \ref{Tab-XSRet-Topic} here.}]$$

The results in Table \ref{Tab-XSRet-Topic} show that the residual joint news coverage (\textit{JointNews}$^\textit{Resid}$) has a significantly negative coefficient, similar in magnitude to the coefficient on \textit{JointNews} in Panel B, Column 1 of Table \ref{Tab-XSRet}. This finding suggests that the predictive power of \textit{JointNews} is unlikely to be driven by the joint coverage of fundamental topic-specific content.\footnote{The predictive power of topic-specific joint news does not have a consistent pattern across topics: 10 topics exhibit a negative association with future stock returns, while 5 show a positive association.} These results provide further confirmation that our findings are conceptually different from those of \cite{Scherbina2015} and \cite{Box2018}, who document that the returns of linked firms positively predict the returns of the focal firm. In contrast, our results are that joint news coverage intensity negatively predicts a firm's one-month-ahead returns.

To summarize, we show that the relationship between joint news coverage, investor attention, and stock returns is unlikely to be driven by fundamental linkages among jointly mentioned stocks or by the coverage of news content belonging to economically important topics. These findings are consistent with the behavioral channels proposed by \cite{BarberOdean2008}, \cite{DaEngelbergGao2011}, and \cite{Barber2021}, which suggest that joint news coverage --- and the associated cross-firm attention spillovers --- trigger higher retail attention, higher contemporaneous valuations, and lower future returns for the covered stocks. \label{subsec_end:XS_EconLink}

\subsection{Joint news and industry returns} \label{subsec:XS_Industry}

We have provided evidence consistent with the first hypothesis, which suggests that joint news coverage triggers investor attention spillovers and leads to high valuations of individual stocks. Next, we examine whether these stock-level effects influence aggregate returns. 

Specifically, we classify each firm based on the 49-industry classification scheme of \citet{Fama1997} and aggregate the firm-level joint news measure to the industry level using value-weighting within each industry as follows:

\begin{equation*}
    \text{JointNews}_{I,t} = \sum_{i \in I} \omega_{it} \text{JointNews}_{it},
\end{equation*}
where $\omega_{it}$ is the value weight of firm $i$ in industry $I$ at time $t$, and $I$ denotes one of the 49 Fama-French industries. We then regress the 49 industry portfolio returns on the industry-level joint news coverage, including industry-level controls and industry- and month-fixed effects, as follows:
\begin{equation}
    R_{I,t+1} = \alpha + \beta \cdot \text{JointNews}_{I,t} + \gamma \cdot X_{I,t} + \varepsilon_{I,t+1}.
\end{equation}

$$[\text{Insert Table \ref{Tab-XSRet-Industry} here.}]$$

Table \ref{Tab-XSRet-Industry} summarizes the results, which show that \textit{JointNews} negatively and significantly predicts next-month industry returns across all specifications. For example, as shown in column (7), a one-standard-deviation increase in \textit{JointNews} predicts a 9.2 basis point decrease in industry returns, on average, after controlling for other attention proxies and two-way fixed effects by industry and by month. This magnitude is comparable to the firm-level regression results, indicating that \textit{JointNews} retains strong predictive power for returns even at the industry level.

In contrast, aggregated self news is insignificant, consistent with the notion that self news is more likely to be associated with idiosyncratic return components, which may be partially diversified away when aggregated to the industry level. \label{subsec_end:XS_Industry}

\section{Time-series Analysis} \label{SecTS}
		
In this section, we build on the stock- and industry-level analysis and examine the ability of JointNews$^M$ to predict market returns. We first examine the baseline in-sample forecasting performance and the impact of market uncertainty and frictions on the forecasting performance of JointNews$^M$. Then, we compare the forecasting performance of JointNews$^M$ with alternative predictors established in previous studies. Next, we analyze the out-of-sample forecasting performance and further examine the predictability by news topics and over longer horizons. We then assess the economic value from an asset allocation perspective. Finally, we provide an identification test using instrumental variables. Panel B of Table \ref{Tab-SummaryStats} and Appendix Table \ref{Tab-Corr} report the summary statistics and the correlation coefficients of the predictors used in forecasting market returns, respectively.

\subsection{Joint news and market returns} \label{subsec:TS_Insample}
		
To investigate the time-series predictability of market returns, we consider the standard univariate predictive regression model,
\begin{equation}\label{eqIS}
    R_{m,t+1} = \alpha + \beta \cdot \textit{X}_{t} +\epsilon_{t+1},
\end{equation}
where $R_{m,t+1}$ is the market excess return, that is, the monthly return on the S\&P500 index in excess of the risk-free rate,\footnote{In unreported analyses, we replicate our key findings by replacing the S\&P500 index returns with the CRSP value-weighted index returns and find that the results remain consistent.} and $X_t$ is one of the return predictors listed in subsection \ref{subsec:Data_other}, specifically, JointNews$^M$, news tone, and investor sentiment proxies, as well as economic predictors \citep{Goyal2008}.

$$[\text{Insert Table \ref{Tab-InSample} here.}]$$
		
Table \ref{Tab-InSample} reports the in-sample forecasting performance. For comparison, all predictors are standardized to have a mean of zero and a standard deviation of one. Both value- and equal-weighted joint news indices, JointNews$^M$ and JointNews$^M_{ew}$, show strong in-sample return predictability, with the coefficient significant at the 1\% level. We illustrate the economic magnitude of the findings using the coefficient estimate of JointNews$^M$. The coefficient of $-0.578$ suggests that a one standard deviation increase in aggregate joint news coverage predicts a substantially lower market return for the next month, by 57.8 bps. In contrast, SelfNews$^{M}$ has no significant forecasting power, suggesting that aggregate self news coverage is not significantly associated with market returns. 

We next investigate whether the predictability of joint news coverage varies across business cycles. Following \citet{Rapach2010}, we compute the $R^2$ statistics separately during economic expansions ($R^2_{up}$) and recessions ($R^2_{down}$),
\begin{equation}\label{eqreg2}
	R^2_c = 1 - \dfrac{\sum^T_{t = 1} 1_{\{t \in \mathbb{T}_c\}} \cdot \epsilon_{t}^2} {\sum^T_{t = 1} 1_{\{t \in \mathbb{T}_c\}} \cdot (R_{m,t} - \bar{R}^m)^2}, \quad c \in \lbrace up, down \rbrace,
\end{equation}
where $1_{\{t \in \mathbb{T}_{up}\}}$ ($1_{\{t \in \mathbb{T}_{up}\}}$) is an indicator that takes a value of one when month $t$ is based on an expansion (recession) period set by the National Bureau of Economic Research (NBER), that is, $\mathbb{T}_{up}$ ($\mathbb{T}_{down}$), and zero otherwise; $\epsilon_{t}$ is the fitted residual, based on the in-sample estimates of the predictive regression model in (\ref{eqIS}); $\bar{R}^m$ is the full-sample mean of $R_{m,t}$; and $T$ is the number of observations for the full sample. The last two columns of Table \ref{Tab-InSample} show that JointNews$^M$ has an $R^2_{up}$ of 1.13\% and an $R^2_{down}$ of 2.73\%.

In summary, these results suggest that, consistent with our hypothesis, the aggregate joint news coverage measures strongly predict lower future market returns, and this predictability is robust across business cycles. \label{subsec_end:TS_Insample}

\subsection{Comparison with alternative predictors} \label{subsec:TS_HorseRace}

In this subsection, we examine whether the forecasting power of the \text{JointNews$^M$} is driven by omitted variables related to business cycle fundamentals, changes in news tones, or investor sentiment. Specifically, we examine the effect of self-news, news tones, investor sentiment, alternative attention proxies, economic fundamentals, and attention-grabbing events (i.e., IPO and M\&A waves).

We estimate the following bivariate predictive regression:
\begin{equation}\label{eqHR}
    R_{m,t+1} = \alpha + \beta \cdot \textit{X}_{t} + \phi \cdot Z_t + \epsilon_{t+1},
\end{equation}
where $X_t$ is \text{JointNews$^M$} or \text{JointNews$^M_{ew}$}, and $Z_t$ is one of the alternative predictors listed in subsection \ref{subsec:Data_other}. Our main focus is on the coefficient $\beta$.

$$[\text{Insert Table \ref{Tab-HorseRace} here.}]$$
		
Table \ref{Tab-HorseRace} shows that the coefficients of \text{JointNews$^M$} are negative and remain statistically significant after controlling for the alternative market return predictors, with magnitudes that are similar to the coefficients reported in Table \ref{Tab-InSample}.  For example, the coefficients of \text{JointNews$^M$} are -1.09 and -0.59, both significant at the 5\% confidence level, after controlling for two powerful predictors that are associated with temporary market overreaction documented in the literature, namely, FEARS by \citet{DaEngelbergGao2015} and market record event indicator (RcrdHigh) by \citet{Yuan2015}, respectively. In addition, the predictability of \text{JointNews$^M$} also survives after controlling for IPO and M\&A wave proxies \citep{Baker2002, Shleifer2003}. The results for \text{JointNews$^M_{ew}$} are similar. Together, these results suggest that JointNews$^M$ contains important information in predicting future market returns that cannot be explained by economic fundamentals, investor sentiment, and attention-grabbing events established in previous studies.

We also control for additional news-based return predictors proposed by recent studies. \cite{Calomiris2019} classify the context and content of news articles into five topics (i.e., markets, governments, commodities, corporate governance and structure, and the extension of credit). The paper then proposes the following word flow measures: entropy, number of articles per month (ArtCount), and the sentiment and frequency for each topic (s[Topic] and f[Topic], respectively). \cite{Bybee2021} analyze the full text of \textit{Wall Street Journal} articles and estimate a topic model to summarize news into 180 topic themes. The paper then quantifies news attention allocated to each theme and identifies the following five news attention estimates as the most important: recession, problems, record high, option/VIX, convertible/preferred.\footnote{We obtain the word flow measures from the website of Harry Mamaysky. The topic attention measures are downloaded from the authors' website \cite{Bybee2021}, available at   \href{http://www.structureofnews.com}{http://www.structureofnews.com}.} 

We estimate Equation (\ref{eqHR}) and analyze the robustness of the predictability of \text{JointNews$^{M}$} with $Z$ being the two sets of alternative measures. The results are presented in Table \ref{Tab-HorseRace-Topic}, with Panels A and B controlling for the lagged \cite{Calomiris2019} word flow predictors and the contemporaneous \cite{Bybee2021} topic attention measures, respectively.\footnote{We follow \cite{Bybee2021}, who use contemporaneous topic attention measures to explain economic conditions and asset prices. Hence, the coefficient on \text{JointNews$^{M}$} captures predictability above and beyond the market return fluctuations that can be explained by the contemporaneous topic attention measures.}

$$[\text{Insert Table \ref{Tab-HorseRace-Topic} here.}]$$

The results show that, consistent with \cite{Calomiris2019} and \cite{Bybee2021}, the word flow measures and topic attention measures have significant power in forecasting and explaining market returns. More importantly, \text{JointNews$^{M}$} remains highly significant. For example, in Panel A, \text{JointNews$^{M}$} has a coefficient of $-0.601$ ($t$-statistic $-2.66$), which is as large as the coefficient in Table \ref{Tab-InSample}. The in-sample adjusted $R^2$ increases by 1.29\% after including \text{JointNews$^{M}$}, which is also comparable to the in-sample $R^2$ in Table \ref{Tab-InSample}. \label{Bybee_Discussion} Similarly, in Panel B, \text{JointNews$^{M}$} has a significantly negative coefficient of $-0.34$ ($t$-statistic $-2.14$) and is associated with a meaningful incremental $R^2$ of 0.36\% even after controlling for the powerful topic-specific attention measures. These findings indicate that \text{JointNews$^{M}$}, constructed by aggregating firm-level joint news coverage, contains relevant information about future market returns that is distinctively different from the latest news-based measures. \label{subsec_end:TS_HorseRace}

\subsection{Out-of-sample forecasts} \label{subsec:TS_OOS}
		
The in-sample analysis utilizes all the available data, thereby allowing for more efficient parameter estimates and more precise return forecasts. However, \citet{Goyal2008}, among others, argue that out-of-sample tests are more relevant for assessing genuine return predictability in real-time and avoiding overfitting issues. In addition, the out-of-sample analyses are much less affected by the finite sample bias, such as the Stambaugh bias \citep{Busetti2013}. Therefore, it is essential to show the out-of-sample predictive performance of \text{JointNews$^M$}.
		
For out-of-sample forecasts at time $t$, we only use the information available up to $t$ to forecast stock returns at $t+1$. Following \citet{Goyal2008}, \citet{Kelly2013}, and many others, we conduct the out-of-sample analysis by recursively estimating the predictive regression (\ref{eqOOS}):
\begin{equation}\label{eqOOS}
    \widehat{R}^m_{t+1} = \widehat{\alpha}_t + \widehat{\beta}_t \cdot \textit{X}_{1:t;t},
\end{equation}
where $X_{1:t;t}$ is the recursively calculated \text{JointNews$^M$} or other return predictors, and $\widehat{\alpha}_t$ and $\widehat{\beta}_t$ are the OLS estimates from regressing $\{R^m_{r+1}\}^{t-1}_{r=1}$ with model (\ref{eqIS}) recursively.\footnote{We impose the theoretically motivated restrictions recommended  by \citet{Campbell2008} when running the predictive regression models.} For comparison, we also carry out the out-of-sample regressions for the alternative predictors used in the prior literature. 

More specifically, to assess the out-of-sample performance, we apply the $R^2_{OS}$ statistic proposed by \citet{Campbell2008}, which measures the proportional reduction in the mean squared forecast error (MSFE) for the predictive regression forecast relative to the historical average benchmark. \citet{Goyal2008} show that the historical average is a very stringent out-of-sample benchmark, and typically, individual economic variables fail to outperform the historical average. To compute $R^2_{OS}$, let $r$ be a fixed number chosen for the initial sample training; this will ensure that the future expected returns can be estimated at time $t = r+1, r+2, \cdots, T$. Subsequently, we compute the following $s=T-r$ out-of-sample forecasts: $\{\widehat{R}^m_{t+1}\}^{T-1}_{t=r}$. Specifically, we use the first 10 years' data from July 1996 through June 2006 as the initial estimation period, and the forecast evaluation period spans from July 2006 through December 2019.
\begin{equation}\label{eqR2oos}
    \widehat{R}^2_{OS} = 1 - \frac {\sum_{t=r}^{T-1}(R_{m,t+1} - \widehat{R}^m_{t+1})^2}{\sum_{t=r}^{T-1}(R_{m,t+1} - \bar{R}^m_{t+1})^2},
\end{equation}
where $\bar{R}^m_{t+1}$ denotes the historical average benchmark corresponding to the constant expected return model ($R_{m,t+1} = \alpha + \epsilon_{t+1}$), that is,
\begin{equation}
    \bar{R}^m_{t + 1} = \dfrac{1}{t} \sum\limits_{s = 1}^t R_{m,s}.
\end{equation}
By construction, the $R^2_{OS}$ statistic lies in the range ($-\infty, 1$]. If $R^2_{OS}>0$, then it would mean that the forecast $\widehat{R}^m_{t+1}$ outperforms the historical average $\bar{R}^m_{t+1}$ in terms of MSFE.\footnote{Notably, the in-sample ($R^2$) uses the full sample mean in the denominator, whereas the out-of-sample ($R^2_{OS}$) is calculated using the historical average return estimated through period $t-1$. This means that the out-of-sample ($R^2$) evaluates the predictive power of the model relative to a benchmark model, such as the historical mean. As a result, it is quite common for the out-of-sample ($R^2$) to be larger than the in-sample ($R^2$). For instance, \cite{RapachRinggenbergZhou2016} used the same approach to evaluate the in- and out-of-sample performance of aggregate short interest and found an in-sample ($R^2$) of 1.24\% and an out-of-sample ($R^2$) of 1.94\%.}

To evaluate the statistical significance of the $R^2_{OS}$, we adopt the MSFE-adjusted statistic of \citet{Clark2007} (CW-test) and the \citet{Diebold1995} statistic modified by \citet{McCracken2007} (DM-test). The CW test tests the null hypothesis that the historical average MSFE is greater than the predictive regression forecast MSFE against the one-sided (right-tail) alternative hypothesis that the historical average MSFE is not greater than the predictive regression forecast MSFE, corresponding to $H_0: R^2_{OS} \leq 0$ against $H_1: R^2_{OS} > 0$. \citet{Clark2007} show that the test has a standard normal limiting distribution when comparing forecasts from the nested models. The DM test examines the null hypothesis that the MSFE of one forecast is equal to that of the other forecast. \citet{McCracken2007} shows that the modified DM test statistic follows a nonstandard normal distribution when testing nested models, and provides bootstrapped critical values for this nonstandard distribution. We expect the benchmark model's MSFE to be smaller than that of the predictive regression model under the null hypothesis. The MSFE-adjusted statistic accounts for the negative expected difference between the historical average MSFE and the predictive regression MSFE under the null hypothesis to ensure that it can reject the null hypothesis even if the $R^2_{OS}$ statistic is negative.

To quantify the economic value of the out-of-sample predictability from the portfolio management perspective, following \citet{Marquering2004, Campbell2008, Goyal2008}, we also calculate realized utility gains for a mean-variance investor on a real-time basis. More specifically, We compute the certainty equivalent return (CER) gain and the Sharpe ratio by considering a mean-variance investor with relative risk aversion parameter $\gamma$ who makes asset allocation decisions across equities and risk-free bills using the out-of-sample predictive regression forecasts \citep[see, e.g.,][]{Kandel1996, Campbell2008, Ferreira2011}. 

At the end of each month $t$, the investor optimally allocates 
\begin{align} \label{eqAlloc}
    w_t = \dfrac{1}{\gamma} \cdot \dfrac{\widehat{R}_{t+1}}{\widehat{\sigma}_{t+1}^2}
\end{align}
of the portfolio to stocks during next month, where $\gamma$ is the degree of risk aversion, $\widehat{R}_{t+1}$ is the out-of-sample forecast of the excess market return, and $\widehat{\sigma}_{t+1}^2$ is the forecast of its variance. The investor then allocates $1-w_t$ of the portfolio to risk-free bills, and the $t+1$ realized portfolio return is
\begin{align}
    R_{t+1}^{p} = w_t R_{t+1} + R_{t+1}^{f},
\end{align}
where $R_{t+1}^{f}$ is the risk-free return. Following \citet{Campbell2008}, we estimate $\widehat{\sigma}_{t+1}$ using a five-year rolling window of past returns and restrict $w_t$ to lie between 0 and 1.5 to exclude short sales and to allow for, at most, 50\% leverage. 

The certainty equivalent return (CER) of the portfolio is
\begin{align}
    \text{CER}_{p} = \widehat{\mu}_{p} - 0.5\widehat{\sigma}_{p}^{2}.
\end{align}
The CER gain ($\Delta$CER) is the difference between the CER for the investor who uses a predictive regression forecast of market return generated by Equation (\ref{eqOOS}) and the CER for an investor who uses the historical average forecast. We multiply this difference by 12 so that it can be interpreted as the annual portfolio management fee that an investor would be willing to pay to have access to the predictive regression forecast instead of the historical average forecast. Following \cite{Rapach2010}, we report
results for $\gamma$ = 3; the results are qualitatively similar for other reasonable $\gamma$ values.

To assess the statistical significance of the CER gain, we apply the testing procedure in \citet{DeMiguel2009} to examine whether the CER gain is indistinguishable from zero. Additionally, we calculate the monthly Sharpe ratio of the portfolio, which is the mean portfolio return in excess of the risk-free rate, divided by the standard deviation of the excess portfolio return. We then test whether the Sharpe ratio of the portfolio strategy based on predictive regression is statistically different from that of the portfolio strategy based on the historical average.

$$[\text{Insert Table \ref{Tab-OOS} here.}]$$

The corresponding results are summarized in Table \ref{Tab-OOS}, which presents three sets of notable findings. First, Panel A shows that both JointNews measures generate positive and significant $R^2_{OS}$s and deliver lower MSFEs than the historical average. Specifically, \text{JointNews$^M$} delivers an $R_{OS}^2$ of 2.9\% and \text{JointNews$^M_{ew}$} delivers an $R_{OS}^2$ of 1.61\%. The strong out-of-sample predictability of \text{JointNews$^M$} for market returns is consistent with our in-sample results in Table \ref{Tab-InSample}. Compared to \text{JointNews$^M$}, all the other predictors (except for Attn$^{PLS}$) show much weaker out-of-sample predictability for excess market returns as shown in Panels B and C.\footnote{Both Attn$^{PLS}$ and ASVI series are available up to December 2017. When we set our out-of-sample evaluation period as July 2006 through December 2017, the $R_{OS}^2$ of \text{JointNews$^M$} is 3.71\%, which is larger than that of Attn$^{PLS}$.} In general, most of the alternative predictors have negative $R_{OS}^2$'s, and their CW- and DM-test statistics are statistically insignificant. 

Second, the fifth and sixth columns show that the annualized CER gain for \text{JointNews}$^{M}$ at the monthly horizon is consistently positive and economically large, and the investment portfolio based on aggregate investor attention generates sizable Sharpe ratios. More specifically, an investor with a risk aversion of three would be willing to pay an annual portfolio management fee of up to 4.23\% to have access to the predictive regression forecast based on \text{JointNews}$^{M}$ instead of using the historical average forecast. The annualized Sharpe ratio of the portfolio formed based on \text{JointNews}$^{M}$ at monthly horizon is 0.62, which is 15\% larger than the market Sharpe ratio of 0.53 for a buy-and-hold strategy. To summarize, asset allocation based on investor attention spillover could potentially offer substantial investment profits, suggesting significant economic value for mean-variance investors.

Lastly, the last two columns of Table \ref{Tab-OOS} show that the predictability of the \text{JointNews$^M$} is significantly strong and stable over both expansion and recession periods, which is consistent with our previous findings in in-sample regressions. 
In summary, both in-sample and out-of-sample results confirm that \text{JointNews$^M$} is a powerful and reliable predictor of excess market returns, and it consistently outperforms the other traditional market return predictors.\label{subsec_end:TS_OOS}

\subsection{Predictability over longer horizons} \label{subsec:TS_Horizon}

So far, we have established that aggregate joint news coverage predicts market returns over a one-month horizon. In this subsection, we turn our focus to longer horizons to investigate the extent to which the associated mispricing can be persistent.\footnote{Because of the limits of arbitrage, mispricings from investor attention may not be eliminated by arbitrageurs over a short monthly horizon. There is some in-sample evidence on the predictability of investor attention for horizons longer than one month in the literature. For example, using the Dow 52-week high as an investor attention proxy, \citet{Li2012} show that the predictive power of investor attention can exist for a multi-month horizon.}

We perform both in- and out-of-sample analysis as in previous sections, based on the following predictive regression:
\begin{equation} \label{eqHorizon}
    R_{m,t+1:t+h} = \alpha + \beta \cdot \text{JointNews}^{M}_{t} + \epsilon_{t+1:t+h},
\end{equation}
where $R_{m,t+1:t+h}$ is $h$-month ahead excess market return from $t$ to $t+h$, and $h$ takes a value from 2 to 4. To analyze the economic value of return predictability at longer horizons, we follow \cite{RapachRinggenbergZhou2016} and assume that the investor rebalances the portfolio at the same frequency as the forecast horizon. For example, when $h = 3$, the investor uses predictive regression (\ref{eqHorizon}) or historical average forecast of the excess return over the next three months ($h = 3$) at the end of each quarter and applies Equation (\ref{eqAlloc}) to determine the stock weight for the next quarter. The investor follows analogous procedures for bi-month and quad-month return forecasts and rebalancing.

$$[\text{Insert Table \ref{Tab-Horizon} here.}]$$

Panel A of Table \ref{Tab-Horizon} reports the in- and out-of-sample univariate regression results. It shows that \text{JointNews$^{M}$} can significantly predict the long-run excess market returns for up to four months. Along with the results in Tables \ref{Tab-HorseRace} and \ref{Tab-OOS} for the one-month horizon, it is clear that the forecasting power first increases with the horizon, reaching its peak around the three-month horizon, and then declines, both in-sample and out-of-sample. 

Panel B of Table \ref{Tab-Horizon} presents the bivariate predictive regression results that include \text{JointNews$^{M}$} and one of the alternative return predictors. It shows that the predictive power of \text{JointNews$^{M}$} for the three-month horizon excess market return is significant and greater than most of the controlling return predictors, including \text{Attn$^{PLS}$}. For the four-month horizon, \text{JointNews$^{M}$} is weaker but still significant at 5\% after controlling for most of the traditional predictors. Economically, a one standard deviation increase in \text{JointNews$^{M}$} reduces the aggregate stock market returns over the next four months by 0.42\% per month. 

In summary, \text{JointNews$^{M}$} significantly predicts future aggregate stock returns, not only at monthly frequency but also at quarterly horizons, both in and out of sample. The result suggests that the clustered arrival of joint news can generate persistent mispricing that is not immediately arbitraged away.\label{subsec_end:TS_Horizon}

\section{Further Analyses} \label{SecFurther}

The evidence of a negative relationship between joint news coverage and future stock returns, combined with findings based on EDGAR IP accesses and Google searches, supports the hypothesis that joint news coverage triggers attention spillovers, which subsequently drive temporary stock overvaluation. In this section, we investigate whether our findings could be explained by a rational explanation, wherein increased investor attention improves risk sharing and lowers a stock's required rate of return (\citealt{Merton1987}). We also investigate whether our results may be driven by omitted variables by exploring plausibly exogenous variations in joint news coverage.

\subsection{Mispricing, uncertainty and market frictions} \label{subsec:Misp}

Our finding of a negative relation between joint news coverage and future stock returns is consistent with the behavioral hypothesis that joint news results in an attention spillover, triggering a temporary overvaluation and subsequent reversals \citep{BarberOdean2008, DaEngelbergGao2011, Barber2021}. However, as explained in the introduction, the relation could also be consistent with a rational alternative explanation: the attention spillover broadens the investor base of a firm, thereby improving risk sharing and reducing the firm's required rate of return \citep{Merton1987}. 

To gain further insight into which mechanism better explains the predictability of \text{JointNews$^M$}, we conduct two tests. The first test focuses on individual stocks and utilizes a direct measure of mispricing to examine whether stocks subject to a greater degree of joint news coverage are more likely to be overvalued. The second test examines time-series return predictability and explores market frictions that make the mispricing mechanism more relevant.

First, to evaluate whether stocks with greater joint news coverage are more likely to be overvalued, we construct two-way sorted portfolios and examine their degree of mispricing. Each month, firms are first sorted into terciles based on one of the firm characteristics --- size, book-to-market ratio, idiosyncratic volatility, or analyst coverage. Within each tercile, firms are further grouped into high and low joint news coverage categories. We use the mispricing proxy (MISP) developed by \citet{Stambaugh2015} to measure the degree of mispricing, where a higher MISP value indicates greater  overvaluation.\footnote{\citet{Stambaugh2015} construct the firm-level mispricing measure by averaging each stock's rankings across 11 return anomalies: net stock issues, composite equity issues, accruals, net operating assets, asset growth, investment-to-assets, distress, O-score, momentum, gross profitability premium, and return on assets. The MISP mispricing measure can be retrieved from R.~F.
Stambaugh's website: \href{https://finance.wharton.upenn.edu/~stambaug/}{https://finance.wharton.upenn.edu/~stambaug/}. Each firm's mispricing measure, joint news coverage, and firm characteristics are demeaned using their respective time-
series averages.} Table \ref{Tab-Mispricing} reports MISP(H-L), the difference in MISP between the high and low joint news coverage portfolios, both for individual characteristic terciles and as an average across the terciles.

$$[\text{Insert Table \ref{Tab-Mispricing} here.}]$$

The table shows that the MISP differential between the high and low joint news coverage groups, MISP(H-L), is positive across all terciles of key firm characteristics. For example, the mispricing level in firms with more joint news is 0.07, 0.50, and 0.28 higher than that of the firms with less joint news in small-, medium-, and large-size firms, respectively. The column ``averages" reports the average MISP differentials across all three terciles for a given firm characteristic, indicating that all MISP differentials are positive and statistically significant. This evidence indicates that high joint news coverage portfolios tend to be more overvalued than those subject to lower levels of joint news coverage, supporting the behavioral mechanism.

In the second test, we focus on time series analysis of market return predictability and explore settings under which one mechanism may be more relevant than the other. If the predictability of \text{JointNews$^M$} is driven by mispricing, we expect the predictability would be stronger during times when market friction is greater and market uncertainty is high, and therefore, arbitrage is more costly. On the other hand, the rational explanation would suggest that the predictability would be equally strong during both high and low uncertainty and friction periods.

To measure market uncertainty, we consider the CBOE volatility index (VIX), the economic uncertainty index (UNC) of \citet{BaliBrownCaglayan2014}, the treasury implied volatility (TIV) of \citet{ChoiMuellerVedolin2017}, the macro uncertainty index (MU) and the financial uncertainty index (FU) of \citet{JuradoLudvigsonNg2015}, the economic policy uncertainty index (EPU) of \citet{BakerBloomDavis2016}, and the investor disagreement index (DSA) by \citet{Huang2021}. For market frictions, we collect the equal-weighted short interest ratio (EWSI) of \citet{RapachRinggenbergZhou2016} and compute the value-weighted average of bid-ask spreads (BAS) and price delay measure of \citet{Hou2005} across S\&P500 stocks.\footnote{The data we use are obtained from the following sources: the VIX is downloaded from the CBOE website; the UNC is downloaded from the website of Turan Bali; the TIV is downloaded from the website of Andrea Vedolin; the MU and FU are downloaded from the website of Sydney Ludvigson; the EPU is downloaded from the website \href{https://www.policyuncertainty.com/}{policyuncertainty.com}; the DSA is downloaded from the website of Dashan Huang; the EWSI is downloaded from the website of Guofu Zhou.} 

Because all of the proxies will have a common uncertainty or market friction component, following \cite{Baker2006, Baker2007}, we use PCA and simple averaging to extract a common component that is closer to the underlying factor. We estimate the following predictive regression, conditioning on the first principal component or equal-weighted average of market uncertainty or friction measures:
\begin{equation}
	R_{m,t+1} = \alpha + \beta \cdot \text{JointNews}^M_{t} +  \gamma \cdot \text{JointNews}^M_{t} \times Z_t  + \phi \cdot Z_t +\epsilon_{t+1},
\end{equation}
where $Z_t$ is an indicator that equals one if the market uncertainty or friction measure is in the top 25\% quantile and equals zero if the corresponding measure is in the bottom 25\% quantile. The variable of interest is $\gamma$, which captures the difference in predictive power between high and low market uncertainty or friction episodes. 

$$[\text{Insert Table \ref{Tab-MktStates} here.}]$$

The results are reported in Table \ref{Tab-MktStates}. The coefficients reported in the ``Diff" columns are all negative and significant across all specifications. For example, in high PCA-Uncertainty periods, a one standard deviation increase in \text{JointNews$^M$} is associated with a 1.01\% decrease in the subsequent returns, which is 1.05\% higher than that in low PCA-Uncertainty periods. Similarly, a one standard deviation increase in \text{JointNews$^M$} predicts a 1.1\% decrease in the subsequent returns in high PCA-Friction periods, which is 1.06\% higher than that in low PCA-Friction periods. Those results suggest that the negative market return predictability of \text{JointNews$^M$} is more likely to be driven by behavioral mechanisms. \label{subsec_end:Misp}

\subsection{Instrumental variable analysis} \label{subsec:IV}

Our result that \text{JointNews$^M$} negatively predicts future market returns is consistent with the hypothesis that joint news coverage predicts returns via an attention-spillover effect. One concern for this interpretation is that joint news coverage is endogenous, and therefore, the relationship between joint news and future returns may be driven by omitted variables that influence joint news coverage.

To address the above concern, we consider episodes of sensational news that are unrelated to the financial market and use these episodes to generate exogenous shocks to joint news coverage. Specifically, we compile a monthly news ``distraction" series from the average of the daily news pressure index of \citet{Eisensee2007}, which is based on the median number of minutes that U.S. news broadcasts devote to the first three news segments.\footnote{To obtain variations in the news pressure that are unrelated to the stock market, we exclude days with macro announcements (CPI, PPI, FOMC, personal income and outlay, and nonfarm payroll) and exclude months in the crisis period or with extreme market return (the top and bottom 10\%) in the annual distribution.} We then define a ``distraction" indicator ($I_\text{Dist}$) as one for observations that belong to the top tercile for that year. We obtain 57 distraction months for the sample period from January 1996 to December 2019.

Table \ref{Tab-IV}, Panel A shows the average level of \text{JointNews$^M$} during the distraction months and
non-distraction months. For the distraction months, \text{JointNews$^M$} has an average of $-0.242$, and it is significantly lower than the non-distraction months (0.061). The results show that sensational news episodes substantially reduce joint coverage of firms, confirming our premise that these episodes distract media attention and lower joint news coverage. 

We then use sensational news as an instrument for \text{JointNews$^M$}. Table \ref{Tab-IV}, Panel B presents the two-stage least squares (2SLS) results. In the first stage, we regress \text{JointNews$^M$} on $I_\text{Dist}$ and find a significantly negative coefficient on $I_\text{Dist}$.
In the second stage, we regress market returns on the instrumented \text{JointNews$^M$}. The results show that the instrumented JointNews$^M$ negatively predicts future market returns, with a significant coefficient of $-3.47$ ($t$-statistic $-2.08$). Economically, a one standard deviation increase in the predicted \text{JointNews$^M$} (12.3\%) lowers the following month's market return by 43 bps, a magnitude that is economically meaningful and is 74\% of that of \text{JointNews$^M$} in Table \ref{Tab-InSample}. 

We acknowledge that our methodology, particularly the use of sensational news as an instrument, has inherent limitations. Sensational news events may not only distract journalists but also directly influence investor sentiment or market conditions, potentially weakening the exclusion restriction. While we attempt to mitigate this concern through additional controls, our findings should be interpreted as suggestive rather than establishing strict causality.

Moreover, our instrument is weak, as indicated by a first-stage $F$-statistic below the conventional threshold of 10. To address potential estimation biases arising from weak instruments \citep[see, e.g.,][]{StaigerStock1997, AndrewsStockSun2019}, we employ the Anderson-Rubin (AR) test \citep[see, e.g.,][]{AndersonRubin1949, StaigerStock1997, Kleibergen2009}, a robust inference method in instrumental variables regression.\footnote{The readers may refer to Section 5.1 of \cite{AndrewsStockSun2019} for a detailed description of the Anderson-Rubin testing procedure. The AR test is automatically reported in Stata using the command \texttt{ivreg2} for IV regression. Unlike traditional approaches that can yield misleading results under weak instrument conditions, the AR test maintains validity by jointly assessing the null hypothesis that the coefficient of the endogenous regressor equals a specific value and that the instruments are exogenous. This ensures correct test size and reliable inference irrespective of instrument strength.} In our analysis, the AR test produced an $F$-statistic of 5.78 with a $p$-value of 0.0169, achieving significance at the 5\% level. This result suggests that our second-stage regression coefficient is statistically significant, even in the presence of a weak instrumental variable. \label{subsec_end:IV}

\section{Conclusions} \label{SecConcl}

We provide novel evidence of news-driven investor attention spillover in financial markets. We find that joint news coverage of firms is positively associated with their contemporaneous stock returns but negatively predicts their future stock returns. A one-standard-deviation increase in joint news coverage is associated with a notable reduction in firms' next-month returns, by 10.3 basis points. Such coverage is also correlated with heightened retail attention, as measured by increased Google search activity. Using SEC EDGAR visit data from unique IP addresses, we provide direct, user-level evidence that the arrival of joint news triggers attention spillovers and high valuations that are transitory in nature.
 
Importantly, the strong firm-level evidence aggregates to predict industry and marketwide returns. A one-standard-deviation increase in aggregate joint news coverage corresponds to a significant decline in market returns, reducing next-month returns by 57.8 basis points. These results remain robust after controlling for a wide range of alternative return predictors, both in-sample and out-of-sample, across expansion and recession periods, and persist for over four months. Notably, this predictability translates into economically significant gains for mean-variance investors in asset allocation.

Additionally, we find that stocks subject to greater joint news coverage tend to be overvalued, with the predictability of market returns being stronger when arbitrage costs are high. These findings are consistent with the hypothesis that the negative association between joint news coverage and future returns is driven by attention-driven overvaluation.  Importantly, we demonstrate that our results are not driven by fundamental linkages across firms or joint news coverage that explicitly feature such linkages, highlighting a novel mechanism that contributes to the existing literature. 

The news media serves a critical role as an information intermediary, conveying both the state of the economy and firm fundamentals to investors. However, it is also prone to distortions, influenced by factors such as consumer preferences, human biases, erroneous inferences, and speculation --- particularly in the age of social media and AI. Our paper contributes to the rapidly growing literature examining how news media shapes investor attention, behavior, and financial market outcomes (e.g., \citealt{Mullainathan2005, Gentzkow2006, cookson2024socialmedia, hirshleifer2025news}). Admittedly, the channel we explore represents just one of many, and further research is essential to advance our understanding of this fast-growing landscape. 

\clearpage

% Bibliography.
\nocite{}
\bibliographystyle{jfe}
\bibliography{MediaBib}
\clearpage

% Results
\clearpage

\clearpage

\begin{table}[htp!]
    \caption{Summary Statistics} \label{Tab-SummaryStats}	
	
    {\small\parindent=2em This table reports summary statistics for all the variables. Panel A reports the summary statistics of the variables used in monthly firm-level panel regressions. We compute the temporal average of each variable first and then report its cross-sectional average (Mean), standard deviation (Stdev), skewness (Skew), 25th percentile (Q25), median (Med), and 75th percentile (Q75). Panel B reports the summary statistics of the variables used in market-level time series regressions. We compute the average (Mean), standard deviation (Stdev), skewness (Skew), kurtosis (Kurt), minimum (Min), maximum (Max), and first-order autocorrelation ($\rho(1)$) of the time series variables. The variable definitions are given in Table \ref{Tab-VarDef}. The sample period spans from July 1996 to December 2019.}
    \smallskip

    \small
    \renewcommand{\arraystretch}{1}
    \begin{tabularx}{\linewidth}{@{\extracolsep{\fill}}lS[table-format = 2.3]S[table-format = 1.3]S[table-format = 2.3]S[table-format = 2.3]S[table-format = 2.3]S[table-format = 2.3]}
        \multicolumn{7}{@{\extracolsep{\fill}}l}{\textbf{Panel A: Firm-level Variables}} \\
        \toprule
        Variable  & {Mean}  & {Stdev}  & {Skew}  & {Q25}  & {Med}   & {Q75} \\
        \midrule
        Return & 0.010 & 0.031 & -3.377 & 0.006 & 0.012 & 0.019 \\
        JointNews$_i$ & 0.066 & 0.076 & 2.777 & 0.016 & 0.047 & 0.087 \\
        SelfNews$_i$ & 0.031 & 0.147 & 3.987 & -0.028 & 0.000 & 0.055 \\
        ASV   & 5.056 & 6.334 & 1.122 & 1.173 & 4.065 & 7.593 \\
        ln(NewIP) & 5.140 & 0.989 & 0.257 & 4.564 & 5.062 & 5.691 \\
        ln(NewIP$_c$) & 2.382 & 1.417 & 0.182 & 1.354 & 2.347 & 3.413 \\
        ln(Size) & 14.216 & 2.217 & -0.563 & 12.813 & 14.657 & 15.724 \\
        ln(BM) & -0.876 & 0.830 & -0.536 & -1.337 & -0.803 & -0.361 \\
        IVOL  & 9.845 & 4.443 & 1.308 & 6.543 & 8.656 & 12.052 \\
        AbnTnvr & 1.755 & 2.866 & 28.569 & 0.836 & 1.362 & 2.135 \\
        Adexp/Sales & 0.032 & 0.048 & 5.145 & 0.008 & 0.016 & 0.038 \\
        ln(1+Analyst) & 1.753 & 0.547 & -0.151 & 1.349 & 1.794 & 2.164 \\
        \bottomrule
    \end{tabularx}
\end{table}%
\clearpage

\begin{table}[htp!]
    \caption*{Table \ref{Tab-SummaryStats} (Cont'd) \quad Summary Statistics}
    
    \small
    \renewcommand{\arraystretch}{1}
    \begin{tabularx}{\linewidth}{@{\extracolsep{\fill}}lS[table-format = 2.3]S[table-format = 1.3]S[table-format = 2.3]S[table-format = 2.3]S[table-format = 2.3]S[table-format = 2.3]S[table-format = 2.3]}
        \multicolumn{8}{@{\extracolsep{\fill}}l}{\textbf{Panel B: Market-level Variables}} \\
	\toprule
	Variables & {Mean}  & {Stdev} & {Skew} & {Kurt} & {Min} & {Max} & {$\rho(1)$}  \\
	\midrule
	\multicolumn{8}{@{\extracolsep{\fill}}l}{\textbf{Returns}} \\
        \quad $R_m$    & 0.007 & 0.043 & -0.658 & 4.051 & -0.169 & 0.035 & 0.042 \\
        \quad $R_f$    & 0.002 & 0.002 & 0.519 & 1.737 & 0.000 & 0.005 & 0.985 \\
        \multicolumn{8}{@{\extracolsep{\fill}}l}{\textbf{News-based Measures}} \\
        \quad JointNews$^M$ & 0.100 & 0.123 & 0.692 & 4.341 & -0.252 & 0.538 & 0.364 \\
        \quad JointNews$^M_{ew}$ & 0.042 & 0.048 & 0.786 & 5.548 & -0.132 & 0.236 & 0.304 \\
        \quad SelfNews$^M$ & -0.003 & 0.177 & -0.013 & 3.351 & -0.541 & 0.531 & 0.019 \\
        \multicolumn{8}{@{\extracolsep{\fill}}l}{\textbf{News Tone, Investor Sentiment and Attention}} \\
        \quad News Tone & -0.072 & 0.054 & 1.098 & 4.033 & -0.184 & 0.105 & 0.827 \\
        \quad Sent$^{BW}$ & 0.093 & 0.655 & 1.951 & 8.145 & -0.961 & 3.140 & 0.964 \\
        \quad Sent$^{PLS}$ & -0.134 & 0.744 & 2.041 & 7.085 & -1.209 & 2.940 & 0.965 \\
        \quad FEARS & 0.002 & 0.037 & 0.009 & 5.564 & -0.139 & 0.116 & -0.259 \\
        \quad Attn$^{PLS}$ & 0.044 & 0.263 & 0.628 & 3.147 & -0.630 & 1.045 & 0.798 \\
        \quad ASVI & -0.143 & 0.226 & 3.556 & 23.094 & -0.475 & 1.508 & 0.381 \\
        \quad RcrdHigh & 0.900 & 0.103 & -1.429 & 4.674 & 0.499 & 1.000 & 0.946 \\
        \multicolumn{8}{@{\extracolsep{\fill}}l}{\textbf{Economic Predictors}} \\
        \quad DP    & -4.015 & 0.202 & -0.068 & 4.293 & -4.524 & -3.281 & 0.976 \\
        \quad DY    & -4.010 & 0.202 & -0.195 & 4.142 & -4.531 & -3.295 & 0.976 \\
        \quad EP    & -3.155 & 0.368 & -2.133 & 9.426 & -4.836 & -2.566 & 0.976 \\
        \quad DE    & -0.860 & 0.422 & 3.350 & 16.006 & -1.244 & 1.379 & 0.983 \\
        \quad SVAR  & -0.293 & 0.500 & -6.669 & 62.931 & -5.809 & -0.015 & 0.699 \\
        \quad BM    & 0.266 & 0.071 & -0.200 & 2.236 & 0.121 & 0.441 & 0.962 \\
        \quad NTIS  & 0.058 & 1.926 & 0.576 & 2.819 & -3.122 & 5.763 & 0.974 \\
        \quad TBL   & -2.102 & 2.008 & -0.519 & 1.722 & -6.170 & -0.010 & 0.996 \\
        \quad LTY   & -4.288 & 1.448 & -0.020 & 1.909 & -7.260 & -1.630 & 0.986 \\
        \quad LTR   & 0.621 & 3.008 & 0.079 & 5.170 & -11.240 & 14.430 & -0.012 \\
        \quad TMS   & 2.186 & 1.322 & -0.051 & 1.929 & -0.410 & 4.530 & 0.977 \\
        \quad DFY   & 0.994 & 0.414 & 3.069 & 15.376 & 0.550 & 3.380 & 0.962 \\
        \quad DFR   & 0.027 & 1.768 & -0.377 & 8.764 & -9.750 & 7.370 & 0.018 \\
        \quad INFL  & 0.177 & 0.350 & -0.879 & 7.652 & -1.915 & 1.222 & 0.479 \\
        \multicolumn{8}{@{\extracolsep{\fill}}l}{\textbf{IPO and M\&A Events}} \\
        \quad IPO   & 21.695 & 16.293 & 1.631 & 6.433 & 0.0     & 106   & 0.762 \\
        \quad MA$_n$ & 1026.33 & 208.988 & 0.429 & 2.797 & 537   & 1600  & 0.803 \\
        \quad MA$_v$ & 121.39 & 64.037 & 0.955 & 3.543 & 24.5  & 345.7 & 0.494 \\
        \multicolumn{8}{@{\extracolsep{\fill}}l}{\textbf{Market Uncertainty and Frictions}} \\
        \quad UNC   & -0.182 & 1.882 & 2.161 & 8.461 & -2.567 & 7.555 & 0.980 \\
        \quad TIV   & 6.683 & 1.966 & 0.828 & 3.783 & 3.495 & 14.330 & 0.892 \\
        \quad MU    & 0.638 & 0.093 & 2.418 & 10.442 & 0.530 & 1.088 & 0.985 \\
        \quad FU    & 0.916 & 0.182 & 0.754 & 3.238 & 0.633 & 1.550 & 0.984 \\
        \quad EPU   & 110.064 & 35.534 & 0.882 & 3.249 & 57.203 & 245.127 & 0.803 \\
        \quad DSA   & 0.136 & 0.917 & 0.377 & 2.461 & -1.705 & 2.401 & 0.983 \\
        \quad EWSI  & 0.349 & 1.016 & -0.019 & 2.428 & -1.679 & 2.738 & 0.992 \\
        \quad BAS   & 0.003 & 0.006 & 5.839 & 56.145 & 0.000 & 0.069 & 0.683 \\
        \quad DLY   & -0.509 & 6.427 & -15.903 & 258.580 & -104.791 & 6.855 & -0.005 \\
        \bottomrule
    \end{tabularx}
\end{table}

\begin{table}[htp!]
    \caption{Joint News Coverage and Investor Attention}\label{Tab-EDGARIP}
    
    {\small\parindent=2em This table reports results from contemporaneous monthly panel regressions of investor attention measures on joint news coverage.  The attention measure in columns 1 and 2 is abnormal Google search volume (ASV, in \%). In columns 3 and 4, the attention measure is the natural logarithm of the number of unique new IPs accessing a firm's EDGAR filings (ln(New IP)). Columns 5 and 6 use the log number of unique new IPs that had previously accessed filings of firms connected through joint news coverage (ln(New IP$_{c}$)). $\text{JointNews}_i$ denotes abnormal joint news coverage for stock $i$ in a given month, calculated as the number of news articles jointly mentioning the stock and other stocks during the month, minus the median joint coverage over the past six months. Similarly, $\text{SelfNews}_i$ captures abnormal self-news coverage, defined as the number of news articles that exclusively mention stock $i$ during the month, minus its past six-month median. Control variables include lagged log firm size (ln(Size)), lagged log book-to-market ratio (ln(BM)), lagged idiosyncratic volatility (IVOL), lagged abnormal turnover (AbnTnvr), lagged advertisement expenses to sales ratio (AdExp/Sales), and lagged log analyst coverage (ln(1+Analyst)). All independent variables are standardized, allowing regression coefficients to be interpreted as the impact of a one-standard-deviation change. Firm and month fixed effects are included. The $t$-statistics, reported in brackets, are based on  \cite{Thompson2011} firm and time double-cluster standard errors with 12 lags. ***, **, and * denote statistical significance at the 1\%, 5\%, and 10\% levels, respectively. The sample period for columns 1 and 2 spans from July 2004 to December 2019, while columns 3 through 6 cover the period from January 2003 to June 2017.}
    \smallskip
	
    \small
    \renewcommand*{\arraystretch}{1}
    \begin{tabularx}{\linewidth}{@{\extracolsep{\fill}}lcccccc}
    \toprule
    & \multicolumn{2}{c}{ASV} & \multicolumn{2}{c}{ln(New IP)} & \multicolumn{2}{c}{ln(New IP$_{c}$)} \\ \cmidrule{2-3}\cmidrule{4-5}\cmidrule{6-7}
    Variables & {(1)}   & {(2)}   & {(3)}   & {(4)}   & {(5)} & {(6)} \\
    \midrule
    JointNews$_i$ & 1.038*** & 0.790*** & 0.048*** & 0.034*** & 0.186*** & 0.150*** \\
          & [5.03] & [5.49] & [9.15] & [6.64] & [8.11] & [7.32] \\
    SelfNews$_i$ & 0.194*** & 0.292*** & 0.025*** & 0.030*** & 0.013** & 0.034*** \\
          & [4.29] & [4.93] & [10.24] & [13.38] & [2.06] & [5.57] \\
    $\ln(\text{Size})$ &       & 0.794 &       & 0.218*** &       & 0.386*** \\
          &       & [1.12] &       & [6.64] &       & [6.97] \\
    $\ln(\text{BM})$ &       & 0.178 &       & 0.054*** &       & 0.076*** \\
          &       & [0.98] &       & [5.09] &       & [4.03] \\
    IVOL  &       & -0.346 &       & 0.002 &       & 0.02 \\
          &       & [-1.10] &       & [0.10] &       & [0.52] \\
    AbnTnvr &       & 1.726*** &       & 0.083*** &       & 0.123*** \\
          &       & [3.36] &       & [5.53] &       & [4.11] \\
    AdExp/Sales &       & 0.967* &       & 0.007 &       & -0.006 \\
          &       & [1.71] &       & [0.32] &       & [-0.16] \\
    $\ln(1+\text{Analyst})$ &       & 0.926*** &       & 0.039*** &       & 0.164*** \\
          &       & [3.10] &       & [9.53] &       & [7.92] \\
    Firm FE & Yes   & Yes    & Yes   & Yes   & Yes   & Yes    \\
    Month FE & Yes   & Yes    & Yes   & Yes   & Yes   & Yes    \\
    Observations & 71,040  & 71,040  & 63,892  & 63,892  & 63,892  & 63,892  \\
    Adj.~$R^2$ (\%) & 16.0 & 16.6 & 91.4 & 91.8 & 75.1 & 75.9 \\
    Within~$R^2$ (\%) & 0.6 & 1.3 & 2.0 & 6.9 & 3.5 & 6.9 \\
    \bottomrule
    \end{tabularx}
\end{table}
\clearpage

\begin{table}[htp!]
    \caption{Joint News Coverage and Stock Returns}\label{Tab-XSRet}
	
    {\small\parindent=2em This table reports contemporaneous and predictive panel regression results of stock returns (in basis points) on firm-level abnormal joint news coverage, $\text{JointNews}_i$, across different horizons. Panels A and B present results at the weekly and monthly frequencies, respectively. $\text{JointNews}_i$ is defined as the number of news articles jointly mentioning stock $i$ and other stocks during a given week (week 0) or month (month 0), minus the median joint coverage over the past 26 weeks or six months, respectively.  Likewise, $\text{SelfNews}_i$ denotes abnormal self-news coverage, measured as the number of articles exclusively mentioning stock $i$ during a given week or month, net of its historical median over the same rolling window. The control variables used in the regressions are the same as those in Table \ref{Tab-EDGARIP}. All independent variables are standardized, allowing regression coefficients to be interpreted as the impact of a one-standard-deviation change. Regressions include firm and time (week or month) fixed effects. The $t$-statistics, reported in brackets, are based on  \cite{Thompson2011} firm and time double-cluster standard errors with 12 lags. ***, **, and * denote statistical significance at the 1\%, 5\%, and 10\% levels, respectively. The sample period spans from July 1996 to June 2019. }
    \smallskip
	
    \small
    \renewcommand*{\arraystretch}{1}
    \setlength{\tabcolsep}{2pt}
    \begin{tabularx}{\linewidth}{@{\extracolsep{\fill}}lccc}
        \multicolumn{4}{@{\extracolsep{\fill}}l}{\textbf{Panel A. Weekly Regressions}} \\
        \toprule
        & Week 0 & Week 1 & Week 1--4 \\ \cmidrule{2-4}
        Variables & (1) & (2) & (3) \\ 
        \midrule
        JointNews$_i$ & 2.259** & -1.810* & -5.965*** \\
          & [1.98] & [-1.95] & [-3.47] \\
        SelfNews$_i$ & 7.344*** & -0.093 & 1.894 \\
              & [4.52] & [-0.14] & [1.08] \\
        $\ln(\text{Size})$ & -66.05*** & -71.42*** & -297.4*** \\
              & [-8.64] & [-9.62] & [-9.49] \\
        $\ln(\text{BM})$ & 0.185 & -0.031 & 0.041 \\
              & [0.04] & [-0.01] & [0.00] \\
        IVOL  & 22.87*** & 7.575 & 25.47 \\
              & [5.18] & [1.53] & [1.54] \\
        AbnTnvr & 0.923 & 0.963* & 5.298** \\
              & [1.58] & [1.76] & [2.02] \\
        AdExp/Sales & -4.217 & 0.109 & 6.781 \\
              & [-0.91] & [0.03] & [1.01] \\
        $\ln(1+\text{Analyst})$ & -4.129** & -0.33 & 0.276 \\
              & [-2.36] & [-0.25] & [0.10] \\
        Firm FE & Yes   & Yes   & Yes \\
        Week FE & Yes   & Yes   & Yes \\
        Month FE & No    & No    & No \\
        Observations & 1,105,856  & 1,107,807  & 1,105,057  \\
        Adj.~$R^2$ (\%) & 21.2 & 21.1 & 22.5 \\
        Within $R^2$ (\%) & 0.3 & 0.2 & 0.9 \\
        $|$Delta$|$ & 14.64 & 4.55  & 6.72 \\
        \bottomrule
    \end{tabularx}
\end{table}
\clearpage

\begin{table}[htp!]
    \caption*{Table \ref{Tab-XSRet} (Cont'd) \quad Joint News Coverage and Stock Returns}
    \smallskip
	
    \small
    \renewcommand*{\arraystretch}{1}
    \setlength{\tabcolsep}{2pt}
    \begin{tabularx}{\linewidth}{@{\extracolsep{\fill}}lccc}
        \multicolumn{4}{@{\extracolsep{\fill}}l}{\textbf{Panel B. Monthly Regressions}} \\
        \toprule
        & Month 1 & Month 2--3 & Month 4--6 \\ \cmidrule{2-4}
        Variables & (1) & (2) & (3) \\ 
        \midrule
        JointNews$_i$ & -10.29*** & -8.238 & -0.133 \\
          & [-3.11] & [-1.08] & [-0.01] \\
        SelfNews$_i$ & -8.052** & 5.402 & -1.152 \\
              & [-2.48] & [1.64] & [-0.33] \\
        $\ln(\text{Size})$ & -278.3*** & -530.2*** & -801.8*** \\
              & [-6.50] & [-6.61] & [-6.64] \\
        $\ln(\text{BM})$ & -11.19 & -31.15 & -43.29 \\
              & [-0.75] & [-1.02] & [-0.95] \\
        IVOL  & -7.96 & -19.59 & -30.37 \\
              & [-0.65] & [-0.82] & [-0.87] \\
        AbnTnvr & 11.55 & 16.17 & 24.097 \\
              & [1.21] & [1.01] & [1.13] \\
        AdExp/Sales & -22.06 & -15.79 & -48.64 \\
              & [-1.62] & [-0.61] & [-1.40] \\
        $\ln(1+\text{Analyst})$ & -10.55** & -21.93** & -37.01*** \\
              & [-2.15] & [-2.24] & [-2.64] \\
        Firm FE & Yes   & Yes   & Yes \\
        Week FE & No    & No    & No \\
        Month FE & Yes   & Yes   & Yes \\
        Observations & 278,377  & 274,624  & 269,074  \\
        Adj.~$R^2$ (\%) & 19.1 & 22.0 & 23.0 \\
        Within~$R^2$ (\%) & 0.6 & 1.1 & 1.6 \\
        $|$Delta$|$ & 3.26  & 2.44  & 0.02 \\
        \bottomrule
    \end{tabularx}
\end{table}
\clearpage

\begin{table}[htp!]
    \caption{Joint News Coverage and Stock Returns,  Controlling for Peer Returns}\label{Tab-XSRet-Peer}
    
    {\small\parindent=2em This table reports results from monthly panel regressions of one-month-ahead returns (in basis points) on lagged abnormal joint news coverage ($\text{JointNews}_i$), controlling for peer firm returns (CFRet) derived from alternative firm connection types. 
    CFRet is computed following \cite{Ali2020}, based on firm connections identified through shared analyst coverage (ana), customer-supplier relationships (cus), common industry classification (ind), technology links (tech), or geographic proximity (geo). Other control variables are the same as those in Table \ref{Tab-XSRet}.  All independent variables are standardized, allowing regression coefficients to be interpreted as the impact of a one-standard-deviation change. Firm and month fixed effects are included. The $t$-statistics, reported in brackets, are based on  \cite{Thompson2011} firm and time double-cluster standard errors with 12 lags. ***, **, and * denote statistical significance at the 1\%, 5\%, and 10\% levels, respectively. The sample period spans from July 1996 to June 2019. }
    \smallskip
	
    \small
    \renewcommand*{\arraystretch}{1}
    \setlength{\tabcolsep}{2pt}
    \begin{tabularx}{\linewidth}{@{\extracolsep{\fill}}lcccccc}
		\toprule
    Variables & {(1)}   & {(2)}   &  {(3)} &  {(4)} & {(5)}   & {(6)}    \\
    \midrule
    JointNews$_i$ & -10.41*** & -10.29*** & -10.29*** & -10.40*** & -10.28*** & -10.46*** \\
          & [-3.16] & [-3.10] & [-3.12] & [-3.13] & [-3.11] & [-3.17] \\
    SelfNews$_i$ & -8.059** & -8.052** & -8.052** & -8.048** & -8.040** & -8.042** \\
          & [-2.47] & [-2.48] & [-2.48] & [-2.48] & [-2.47] & [-2.48] \\
    $\text{CFRet}^{ana}$ & -0.020 &       &       &       &       & -0.023 \\
          & [-1.39] &       &       &       &       & [-1.46] \\
    $\text{CFRet}^{cus}$ &       & 0.001 &       &       &       & 0.003 \\
          &       & [0.13] &       &       &       & [0.49] \\
    $\text{CFRet}^{ind}$ &       &       & 0.000 &       &       & 0.013 \\
          &       &       & [-0.00] &       &       & [0.83] \\
    $\text{CFRet}^{tech}$ &       &       &       & -0.024 &       & -0.024 \\
          &       &       &       & [-1.11] &       & [-1.25] \\
    $\text{CFRet}^{geo}$ &       &       &       &       & 0.005 & 0.006 \\
          &       &       &       &       & [0.83] & [1.23] \\
    Controls  & Yes   & Yes   & Yes   & Yes   & Yes   & Yes \\
    Firm FE & Yes   & Yes   & Yes   & Yes   & Yes   & Yes \\
    Month FE & Yes   & Yes   & Yes   & Yes   & Yes   & Yes \\
    Observations & 278,377  & 278,377  & 278,377  & 278,377  & 278,377  & 278,377  \\
    Adj.~$R^2$ (\%) & 19.1 & 19.1 & 19.1 & 19.1 & 19.1 & 19.2 \\
    Within~$R^2$ (\%) & 0.65 & 0.64 & 0.64 & 0.65 & 0.64 & 0.65 \\
    \bottomrule
    \end{tabularx}
\end{table}

\begin{table}[htp!]
    \caption{Joint News Coverage and Stock Returns, by News Topics}\label{Tab-XSRet-Topic}
	
    {\small\parindent=2em This table reports results from monthly panel regressions of one-month-ahead returns (in basis points) on lagged abnormal joint news coverage, where joint news coverage is decomposed by news topics. \text{JointNews$^{Topic}$} denotes the abnormal joint news coverage related to a specific topic, with topic classifications based on \cite{Scherbina2015}. The topics include: supply chain (Sply), partnerships (Ptnr), M\&A (MA), parent/subsidiary relations (Prnt), legal (Legal), regulation (Regl), labor, production and infrastructure (Lbr), executive compensation and corporate governance (CGvn), management news (Mgmt), common customer (Cstm), cross-investment (Ivst), natural resources (NRes), energy (Engy), technology (Tech), and geopolitical (Gptc). The residual (Resid) is computed as the regression residual from regressing abnormal joint news coverage across all news (\text{JointNews}) on all topic-specific \text{JointNews$^{Topic}$}.  The control variables are the same as those in Table \ref{Tab-XSRet} and are omitted here for brevity. All independent variables are standardized, allowing regression coefficients to be interpreted as the impact of a one-standard-deviation change. Firm and month fixed effects are included. The $t$-statistics, reported in brackets, are based on  \cite{Thompson2011} firm and time double-cluster standard errors with 12 lags. ***, **, and * denote statistical significance at the 1\%, 5\%, and 10\% levels, respectively. The sample period spans from July 1996 to June 2019.}
    \medskip
	
    \small
    \setlength{\tabcolsep}{2pt}
    \begin{tabularx}{\linewidth}{@{\extracolsep{\fill}}lcccccccc}
    \toprule
    Variables & Resid & Sply & Ptnr & MA  & Prnt & Legal & Regl & Lbr \\
    \midrule
    JointNews$^\textit{Topic}_i$ & -17.56*** & -5.990 & -9.144 & 0.813 & -63.78 & -9.177 & -8.446 & -13.14 \\
          & [-2.75] & [-0.99] & [-1.39] & [0.35] & [-0.95] & [-1.41] & [-1.29] & [-0.70] \\
    SelfNews$_i$ & -8.073** & -8.093** & -8.105** & -8.098** & -8.097** & -8.064** & -8.087** & -8.082** \\
          & [-2.49] & [-2.49] & [-2.50] & [-2.49] & [-2.49] & [-2.48] & [-2.49] & [-2.48] \\
    Controls & Yes   & Yes   & Yes   & Yes   & Yes   & Yes   & Yes   & Yes \\
    Firm FE & Yes   & Yes   & Yes   & Yes   & Yes   & Yes   & Yes   & Yes \\
    Month FE & Yes   & Yes   & Yes   & Yes   & Yes   & Yes   & Yes   & Yes \\
    Observations & 278,377  & 278,377  & 278,377  & 278,377  & 278,377  & 278,377  & 278,377  & 278,377  \\
    Adj.~$R^2$ (\%) & 19.1 & 19.1 & 19.1 & 19.1 & 19.1\% & 19.1\% & 19.1 & 19.1 \\
    Within~$R^2$ (\%) & 0.64 & 0.63 & 0.63 & 0.63 & 0.63\% & 0.64\% & 0.63 & 0.63 \\
    
    \midrule
          &  CGvn   & Mgmt  & Cstm & Ivst & NRes & Engy & Tech  & Gptc \\
    \midrule
    JointNews$^\textit{Topic}_i$ & -20.74 & 5.009 & -28.99** & 2.474 & 17.10*** & 5.226 & -5.334 & -3.391 \\
          & [-0.34] & [0.12] & [-2.08] & [0.31] & [2.98] & [0.94] & [-0.44] & [-0.75] \\
    SelfNews$_i$ & -8.102** & -8.100** & -8.101** & -8.101** & -8.099** & -8.099** & -8.093** & -8.104** \\
          & [-2.50] & [-2.50] & [-2.50] & [-2.50] & [-2.50] & [-2.49] & [-2.49] & [-2.50] \\
    Controls & Yes   & Yes   & Yes   & Yes   & Yes   & Yes   & Yes   & Yes \\
    Firm FE & Yes   & Yes   & Yes   & Yes   & Yes   & Yes   & Yes   & Yes \\
    Month FE & Yes   & Yes   & Yes   & Yes   & Yes   & Yes   & Yes   & Yes \\
    Observations & 278,377  & 278,377  & 278,377  & 278,377  & 278,377  & 278,377  & 278,377  & 278,377  \\
    Adj.~$R^2$ (\%) & 19.1 & 19.1 & 19.1 & 19.1 & 19.1 & 19.1 & 19.1 & 19.1 \\
    Within~$R^2$ (\%) & 0.63 & 0.63 & 0.63 & 0.63 & 0.64 & 0.63 & 0.63 & 0.63 \\
    \bottomrule   
    \end{tabularx}
\end{table}
\clearpage

\begin{table}[htp!]
    \caption{Joint News Coverage and Industry Returns} \label{Tab-XSRet-Industry}
	
    {\small\parindent=2em This table reports predictive panel regression results of monthly industry returns (in bps) on industry-level abnormal joint news $\text{JointNews}_I$, which is constructed by aggregating the firm-level abnormal joint news with value weights within each industry. The control variables are the same as those in Table \ref{Tab-XSRet}, which are also aggregated to the industry level with value weights. All independent variables are standardized, allowing regression coefficients to be interpreted as the impact of a one-standard-deviation change. Firm and month fixed effects are included. The $t$-statistics, reported in brackets, are based on  \cite{Thompson2011} firm and time double-cluster standard errors with 12 lags.  ***, **, and * denote statistical significance at the 1\%, 5\%, and 10\% levels, respectively. The sample period spans from July 1996 to December 2019.}
    \medskip
	
    \small
    \setlength{\tabcolsep}{2pt}
    \begin{tabularx}{\linewidth}{@{\extracolsep{\fill}}lccccccc}
    \toprule
    & \multicolumn{7}{@{\extracolsep{\fill}}c}{Fama-French 49 Industries} \\ \cmidrule{2-8}
    Variables & (1)   & (2)   & (3)   & (4)   & (5)   & (6)   & (7) \\
    \midrule
    JointNews$_I$ & -10.54* & -12.30** & -12.14** & -12.48** & -12.34** & -11.22** & -9.234** \\
          & [-2.01] & [-2.41] & [-2.49] & [-2.56] & [-2.45] & [-2.28] & [-2.03] \\
    SelfNews$_I$ & -9.413 & -9.258 & -9.410 & -9.290 & -9.244 & -9.928 & -10.10 \\
          & [-1.24] & [-1.21] & [-1.23] & [-1.22] & [-1.21] & [-1.25] & [-1.29] \\
    ln(Size) & -29.24* &       &       &       &       &       & -36.65** \\
          & [-1.69] &       &       &       &       &       & [-2.25] \\
    ln(BM) &       & 5.828 &       &       &       &       & -1.729 \\
          &       & [0.28] &       &       &       &       & [-0.09] \\
    IVOL  &       &       & -12.56 &       &       &       & -17.39 \\
          &       &       & [-0.97] &       &       &       & [-1.48] \\
    AbnTnvr &       &       &       & 2.248 &       &       & -0.026 \\
          &       &       &       & [0.68] &       &       & [-0.00] \\
    AdExp/Sales &       &       &       &       & -0.174 &       & 8.535 \\
          &       &       &       &       & [-0.04] &       & [1.64] \\
    ln(1+Analyst) &       &       &       &       &       & -18.18 & -10.90 \\
          &       &       &       &       &       & [-1.57] & [-1.03] \\
    Industry FE & Yes   & Yes   & Yes   & Yes   & Yes   & Yes   & Yes \\
    Month FE & Yes   & Yes   & Yes   & Yes   & Yes   & Yes   & Yes \\
    Observations & 13,769  & 13,769  & 13,769  & 13,769  & 13,769  & 13,769  & 13,769  \\
    Adj.~$R^2$ (\%) & 0.12 & 0.05 & 0.08 & 0.04 & 0.04 & 0.07 & 0.16 \\
    Within~$R^2$ (\%) & 0.14 & 0.08 & 0.11 & 0.07 & 0.07 & 0.10 & 0.07 \\
    \bottomrule
    \end{tabularx}
\end{table}
\clearpage

\begin{table}[htp!]
    \caption{In-Sample Forecasting of Market Returns}\label{Tab-InSample}
			
    {\small\parindent=2em This table provides in-sample estimation results for the predictive regression
	$$R_{m,t+1} = \alpha + \beta X_t + \epsilon_{t+1},$$
    where $R_{m,t+1}$ denotes the monthly excess market return (in \%) and $X_t$ is one of the return predictors. In Panel A, the return predictor is the value- and the equal-weighted aggregate joint news index (JointNews$^M$ and JointNews$^M_{ew}$) and the value-weighted aggregate self news index (SelfNews$^M$). In Panel B, the return predictor is the news tone maintained by Thomson Reuters, the investor sentiment measures proposed by \citet{Baker2007} (Sent$^{BW}$) and \citet{Huang2015} (Sent$^{PLS}$), the FEARS index proposed by \citet{DaEngelbergGao2015}, the composite investor attention measure proposed by \citet{Chen2020} (Attn$^{PLS}$), the abnormal search volume for the major stock indices used in \citet{LiuPengTang2023} (ASVI), and nearness to the Dow historical high proposed by \citet{Chen2020} (RcrdHigh) based on \citet{Yuan2015}. For Panel C, the return predictors are economic predictors following \cite{Goyal2008}: the log dividend price ratio (DP), log dividend yield (DY), log earnings-price ratio (EP), log dividend-payout ratio (DE), stock return variance (SVAR), book-to-market ratio (BM), net equity expansion (NTIS), Treasury bill rate (TBL), long-term bond yield (LTY), long-term bond return (LTR), term spread (TMS), default yield spread (DFY), default return spread (DFR) and inflation rate (INFL). The $t$-statistics are based on Newey-West standard errors with the six lags. ***, **, and * denote statistical significance at the 1\%, 5\%, and 10\% levels, respectively. The sample period spans from July 1996 to December 2019.}
    \smallskip
	
    \renewcommand*{\arraystretch}{0.92}
    \small
    \begin{tabularx}{\textwidth}{@{\extracolsep{\fill}}lS[table-format = -1.3]S[table-format = -1.3]S[table-format = 1.3]S[table-format = 1.3]S[table-format = 2.3]}
	\toprule
		Predictors & {$\widehat{\beta}$} & {$t$-stat.} & {$R^2$(\%)} & {$R^2_{up}$(\%)} & {$R^2_{down}$(\%)} \\
		\midrule
		\multicolumn{6}{@{\extracolsep{\fill}}l}{\textbf{Panel A: News-based Measures}} \\
        \quad JointNews$^M$ & -0.578** & -2.563 & 1.809 & 1.130 & 2.731 \\
        \quad JointNews$_{ew}^M$ & -0.381** & -2.201 & 0.782 & 0.846 & 0.069 \\
        \quad SelfNews$^M$ & -0.088 & -0.381 & 0.042 & 0.002 & 1.115 \\
        \multicolumn{6}{@{\extracolsep{\fill}}l}{\textbf{Panel B: News Tone, Investor Sentiment and Attention}} \\
        \quad News Tone & -0.334 & -1.171 & 0.606 & 0.119 & 0.345 \\
        \quad Sent$^{BW}$ & -0.556*** & -2.674 & 1.681 & 2.572 & 0.064 \\
        \quad Sent$^{PLS}$ & -0.704*** & -4.266 & 2.694 & 3.487 & 0.235 \\
        \quad FEARS & -0.750 & -0.959 & 2.637 & 3.175 & 21.740 \\
        \quad Attn$^{PLS}$ & -0.641* & -1.849 & 2.209 & 0.647 & 11.465 \\
        \quad ASVI & -0.974*** & -2.982 & 6.136 & 1.961 & 11.493 \\
        \quad RcrdHigh & -0.246 & -0.777 & 0.329 & 0.924 & 6.893 \\
        
        \multicolumn{6}{@{\extracolsep{\fill}}l}{\textbf{Panel C: Economic Predictors}} \\
        \quad DP    & 0.589 & 1.431 & 1.885 & 5.508 & 0.170 \\
        \quad DY    & 0.629* & 1.830 & 2.152 & 4.506 & 1.044 \\
        \quad EP    & 0.237 & 0.525 & 0.305 & 1.467 & 14.079 \\
        \quad DE    & 0.075 & 0.214 & 0.030 & 0.031 & 8.085 \\
        \quad SVAR  & 0.592** & 2.365 & 1.905 & 0.016 & 3.392 \\
        \quad BM    & 0.288 & 1.206 & 0.451 & 0.705 & 0.631 \\
        \quad NTIS  & -0.399 & -1.030 & 0.865 & 0.000 & 2.083 \\
        \quad TBL   & 0.165 & 0.642 & 0.148 & 0.224 & 0.612 \\
        \quad LTY   & 0.339 & 1.204 & 0.618 & 0.609 & 0.015 \\
        \quad LTR   & 0.272 & 1.504 & 0.400 & 0.295 & 0.762 \\
        \quad TMS   & -0.117 & -0.425 & 0.074 & 0.019 & 1.390 \\
        \quad DFY   & -0.297 & -0.801 & 0.480 & 0.000 & 0.092 \\
        \quad DFR   & 0.295 & 1.025 & 0.471 & 0.118 & 2.031 \\
        \quad INFL  & 0.326 & 0.812 & 0.577 & 0.279 & 15.899 \\
	\bottomrule
    \end{tabularx}
\end{table}
\clearpage

\begin{table}[htp!]
    \caption{Horse Race for In-Sample Forecasting of Market Returns}\label{Tab-HorseRace}
	
    {\small\parindent=2em This table provides in-sample estimation results for the bivariate predictive regression of monthly excess market returns on JointNews$^M$ or JointNews$^M_{ew}$, and one of the other predictors, $Z_t$.
	$$R_{m,t+1} = \alpha + \beta X_t + \phi Z_t + \epsilon_{t+1}, $$
    where $R_{m,t+1}$ denotes the monthly excess market return (\%), JointNews$^M$ and JointNews$^M_{ew}$ are the value- and the equal-weighted aggregate joint news indices, and $Z$ is one of the other predictors in Table \ref{Tab-InSample} and the monthly number of IPO cases maintained by Jay Ritter (IPO) and the monthly number and gross value of M\&A cases  (MA$_n$ and MA$_v$) maintained by the IMAA Institute. The significance of the estimates is based on Newey-West $t$-statistics with the 12 lags. ***, **, and * denote statistical significance at the 1\%, 5\%, and 10\% levels, respectively. The sample period spans from July 1996 to December 2019.}
    \smallskip
	
    \renewcommand*{\arraystretch}{1}
    \small      
    \begin{tabularx}{\linewidth}{@{\extracolsep{\fill}}lS[table-format = -1.3,table-space-text-post ={*}]S[table-format = -1.3,table-space-text-post ={*}]S[table-format = 1.3]S[table-format = -1.3,table-space-text-post ={*}]S[table-format = -1.3,table-space-text-post ={*}]S[table-format = 1.3]}
    \toprule
    & \multicolumn{3}{c}{JointNews$^M$} & \multicolumn{3}{c}{JointNews$^M_{ew}$} \\ \cmidrule{2-4}\cmidrule{5-7}
    Predictors & {$\widehat{\beta}$}  & {$\widehat{\phi}$}   & {$R^2$(\%)}    & {$\widehat{\beta}$}  & {$\widehat{\phi}$}   & {$R^2$(\%)} \\
    \midrule
    \multicolumn{7}{@{\extracolsep{\fill}}l}{\textbf{Controlling for Self News}} \\
    \quad SelfNews$^M$ & -0.617*** & -0.205 & 2.028 & -0.407** & -0.156 & 0.911 \\
    \multicolumn{7}{@{\extracolsep{\fill}}l}{\textbf{Controlling for News Tone, Investor Sentiment and Attention}} \\
    \quad News Tone & -0.534** & -0.232 & 2.089 & -0.346** & -0.294 & 1.243 \\
    \quad Sent$^{BW}$ & -0.529** & -0.505** & 3.184 & -0.338** & -0.529** & 2.294 \\
    \quad Sent$^{PLS}$ & -0.513** & -0.654*** & 4.107 & -0.330* & -0.680*** & 3.279 \\
    \quad FEARS & -1.086*** & -0.696 & 9.603 & -0.738*** & -0.740 & 5.075 \\
    \quad Attn$^{PLS}$ & -0.593*** & -0.614* & 3.987 & -0.359** & -0.638* & 2.846 \\
    \quad ASVI & -0.516** & -0.835*** & 8.050 & -0.220 & -0.932*** & 6.511 \\
    \quad RcrdHigh & -0.594** & -0.281 & 2.234 & -0.381** & -0.246 & 1.110 \\     
    \multicolumn{7}{@{\extracolsep{\fill}}l}{\textbf{Controlling for Economic Predictors}} \\
    \quad DP    & -0.553** & 0.565 & 3.540 & -0.381** & 0.589 & 2.668 \\
    \quad DY    & -0.534** & 0.590* & 3.688 & -0.368** & 0.622* & 2.882 \\
    \quad EP    & -0.571** & 0.218 & 2.066 & -0.378** & 0.232 & 1.075 \\
    \quad DE    & -0.579** & 0.080 & 1.844 & -0.382** & 0.079 & 0.816 \\
    \quad SVAR  & -0.469*** & 0.488* & 3.036 & -0.318** & 0.557** & 2.443 \\
    \quad BM    & -0.586** & 0.303 & 2.306 & -0.407** & 0.321 & 1.337 \\
    \quad NTIS  & -0.642*** & -0.483 & 3.056 & -0.448** & -0.464 & 1.926 \\
    \quad TBL   & -0.576** & 0.159 & 1.946 & -0.390** & 0.185 & 0.968 \\
    \quad LTY   & -0.538** & 0.252 & 2.141 & -0.355** & 0.309 & 1.293 \\
    \quad LTR   & -0.593** & 0.301 & 2.297 & -0.389** & 0.283 & 1.215 \\
    \quad TMS   & -0.574** & -0.026 & 1.813 & -0.372** & -0.055 & 0.798 \\
    \quad DFY   & -0.548*** & -0.224 & 2.076 & -0.361** & -0.271 & 1.180 \\
    \quad DFR   & -0.573** & 0.283 & 2.244 & -0.381** & 0.295 & 1.254 \\
    \quad INFL  & -0.547*** & 0.260 & 2.171 & -0.354** & 0.293 & 1.245 \\
    \multicolumn{7}{@{\extracolsep{\fill}}l}{\textbf{Controlling for IPO and M\&A Events}} \\
    \quad IPO   & -0.583*** & 0.212 & 2.053 & -0.389** & 0.213 & 1.028 \\
    \quad MA$_n$ & -0.583*** & 0.212 & 2.053 & -0.389** & 0.213 & 1.028 \\
    \quad MA$_v$ & -0.588** & -0.250 & 2.149 & -0.393** & -0.245 & 1.109 \\
    \bottomrule
    \end{tabularx}	
\end{table}
\clearpage

\begin{table}[htp!]
    % \addtocounter{table}{-1}
    % \renewcommand{\thetable}{\arabic{table}B}
    
    \caption{In-Sample Forecasting of Market Returns Controlling for Textual-based Variables}\label{Tab-HorseRace-Topic}
    
    {\small\parindent=2em This table reports the in-sample predictive regression results of regressing monthly excess market returns on JointNews$^M$ after controlling for variables constructed from textual analysis. Panel A reports the results of controlling for lagged \cite{Calomiris2019} measures for the U.S. market: entropy, number of articles per month (ArtCount), and article sentiment (s[Topic]) and frequency (f[Topic]) for Topic in markets, commodities, corporate sector, credit, and government. Panel B reports the results of controlling for the AR(1) innovations of five contemporaneous news attention measures in \cite{Bybee2021}: Recession, Problems, Record High, Option/VIX, and Convertible/preferred. JointNews$^M$ is the value-weighted aggregate joint news index. The $t$-statistics reported in brackets are based on the Newey-West robust standard errors with 12 lags. ***, ** and * denote statistical significance at the 1\%, 5\%, and 10\% levels, respectively. The sample period spans from July 1996 to December 2019.}
    \smallskip
    
    \renewcommand*{\arraystretch}{1}
    \small
    \begin{tabularx}{\linewidth}{@{\extracolsep{\fill}}lcclcc}
    \toprule
    & \multicolumn{2}{c}{Panel A: } &       & \multicolumn{2}{c}{Panel B: } \\
    & \multicolumn{2}{c}{\cite{Calomiris2019}} &       & \multicolumn{2}{c}{\cite{Bybee2021}} 
    \\ \cmidrule{2-3}\cmidrule{5-6} 
    & (1)   & (2)   &       & (3)   & (4) \\
    \midrule
    JointNews$^M$ &       & -0.603*** & JointNews$^M$ &       & -0.340** \\
          &       & [-2.66] &       &       & [-2.15] \\
    Entropy & 3.032 & 4.942 & Recession & -1.416*** & -1.395*** \\
          & [0.59] & [1.02] &       & [-4.98] & [-4.95] \\
    ArtCount & 1.502** & 1.662** & Problems & -0.976*** & -0.964*** \\
          & [2.13] & [2.42] &       & [-3.03] & [-2.96] \\
    sMkt  & -0.172 & -0.240 & Record High & 0.583** & 0.589** \\
          & [-0.15] & [-0.22] &       & [2.50] & [2.48] \\
    fMkt  & -0.866 & -0.823 & Option/VIX & -0.449* & -0.415* \\
          & [-1.19] & [-1.08] &       & [-1.88] & [-1.67] \\
    sComms & 0.103 & -0.048 & Convertible/preferred & -0.497** & -0.467** \\
          & [0.29] & [-0.12] &       & [-2.23] & [-2.03] \\
    fComms & 0.416 & 0.462 &       &       &  \\
          & [0.86] & [0.97] &       &       &  \\
    sCorp & 0.485 & 0.561 &       &       &  \\
          & [0.52] & [0.62] &       &       &  \\
    fCorp & -0.088 & 0.020 &       &       &  \\
          & [-0.16] & [0.04] &       &       &  \\
    sCredit & 0.407 & 0.349 &       &       &  \\
          & [0.64] & [0.55] &       &       &  \\
    fCredit & -0.400  & -0.395 &       &       &  \\
          & [-0.59] & [-0.59] &       &       &  \\
    sGovt & -0.576 & -0.554 &       &       &  \\
          & [-0.75] & [-0.75] &       &       &  \\
    Intercept & -6.982 & -11.59 & Intercept & 0.436** & 0.490** \\
          & [-0.56] & [-0.99] &       & [1.96] & [2.15] \\
    Observations & 253   & 253   & Observations & 256   & 251 \\
    Adj.~$R^2$ (\%) & -0.07 & 1.30 & Adj.~$R^2$ (\%) & 25.09 & 25.45 \\
    \bottomrule
    \end{tabularx}
    
    {\footnotesize Note: fGovt is omitted due to the collinearity issue in the data.}
\end{table}
\clearpage

\begin{table}[htp!]
	\caption{Out-of-Sample Forecasting of Market Returns}\label{Tab-OOS}
			
    {\small\parindent=2em This table reports the out-of-sample performance of monthly market excess return predictors. Panel A provides the results using the value- and the equal-weighted aggregate joint news index (JointNews$^M$ and JointNews$^M_{ew}$), and the value-weighted aggregate self news index (SelfNews$^M$). Panel B shows results of the News Tone by TRNA, the investor sentiment measures proposed by \citet{Baker2007} (Sent$^{BW}$) and \citet{Huang2015} (Sent$^{PLS}$), the FEARS index proposed by \citet{DaEngelbergGao2015}, the composite investor attention measure proposed by \citet{Chen2020} (Attn$^{PLS}$), the abnormal search volume for the major stock indices used in \citet{LiuPengTang2023} (ASVI), and nearness to the Dow historical high proposed by \citet{Chen2020} (RcrdHigh) based on \citet{Yuan2015}. Panel C shows results using economic predictors \citep{Goyal2008} (see Table \ref{Tab-InSample} for the full list). All the predictors and regression slopes are estimated recursively using the data available at the forecast formation time $t$. $R^2_{OS}$ is the out-of-sample $R^2$ statistic as in \cite{Campbell2008}. $R^2_{OS,up}$ $(R^2_{OS,down})$ statistics are calculated over NBER-dated business-cycle expansions (recessions) based on the unconstrained model. CW test is the \citet{Clark2007} MSFE-adjusted statistic. DM test is the \citet{Diebold1995} statistic for forecast evaluation. $\Delta$CER is the certainty equivalent gain, and SRatio denotes the annualized Sharpe ratios of portfolios formed based on the predictor's forecast. ***, **, and * denote statistical significance at the 1\%, 5\%, and 10\% levels, respectively. The full sample period spans from July 1996 to December 2019, and the out-of-sample evaluation period is from July 2006 to December 2019.}
    \smallskip
	
    \renewcommand*{\arraystretch}{1}
    \small
    \centering
    \begin{tabularx}{\linewidth}{@{\extracolsep{\fill}}lS[table-format = -1.3,table-space-text-post ={**}]S[table-format = -1.3,table-space-text-post ={*}]S[table-format = 1.3]S[table-format = -1.3, table-space-text-post ={**}]S[table-format = 1.3]S[table-format = -1.3,table-space-text-post ={*}]S[table-format = -1.3,table-space-text-post ={*}]}
    \toprule
    Predictors & {$R^2_{OS}$(\%)} & {CW test} & {DM test} & {$\Delta$CER(\%)} & {SRatio} & {$R^2_{OS, up}$(\%)} & {$R^2_{OS,down}$(\%)}  \\
    \midrule
    \multicolumn{8}{@{\extracolsep{\fill}}l}{\textbf{Panel A: News-based Measures}}  \\
    \quad JointNews$^M$ & 2.903*** & 2.848 & 1.463* & 4.225** & 0.624** & 4.170 & 0.608 \\
    \quad JointNews$_{ew}^M$ & 1.613*** & 2.346 & 1.117 & 1.655 & 0.436 & 2.757 & -0.461 \\
    \quad SelfNews$^M$ & -0.614 & -1.799 & -0.880 & -0.531 & 0.272 & -0.633 & -0.580 \\
    \multicolumn{8}{@{\extracolsep{\fill}}l}{\textbf{Panel B: News Tone, Investor Sentiment and Attention}}  \\
    \quad News Tone & -0.259 & 0.468 & -0.236 & 1.583 & 0.436 & -0.912 & 0.925 \\
    \quad Sent$^{BW}$ & 0.638 & 0.745 & 0.413 & -0.372 & 0.329 & 2.964 & -3.576 \\
    \quad Sent$^{PLS}$ & 0.502** & 2.024 & 0.346 & 3.688*** & 0.581*** & 1.501 & -1.306 \\
    \quad FEARS & -4.358 & -2.381 & -2.088 & -5.479** & -0.287 & -5.029 & -3.142 \\
    \quad Attn$^{PLS}$ & 2.660*** & 2.927 & 1.616* & 5.038* & 0.687* & 3.130 & 1.809 \\
    \quad ASVI & -0.033 & 1.042 & -0.020 & 0.058 & 0.314 & 0.821 & -1.580 \\
    \quad RcrdHigh & -0.654 & -0.173 & -0.574 & -2.819* & 0.110 & 0.590 & -2.907 \\
    \multicolumn{8}{@{\extracolsep{\fill}}l}{\textbf{Panel C: Economic Predictors}} \\
    \quad DP    & -4.149 & -0.339 & -1.239 & 0.696 & 0.388 & 4.524 & -19.86 \\
    \quad DY    & -3.030 & -0.115 & -0.990 & 1.147 & 0.419 & 4.764 & -17.15 \\
    \quad EP    & 0.032 & 1.204 & 0.035 & 3.160*** & 0.558*** & -0.152 & 0.363 \\
    \quad DE    & 0.123 & 0.483 & 0.425 & 0.156 & 0.322 & 0.070 & 0.217 \\
    \quad SVAR  & 0.696* & 1.572 & 0.979 & 0.985 & 0.387 & 0.902 & 0.323 \\
    \quad BM    & -0.005 & 0.307 & -0.007 & -0.905 & 0.270 & 1.211 & -2.208 \\
    \quad NTIS  & 0.077 & 1.044 & 0.862 & 0.000 & 0.308 & 0.000 & 0.217 \\
    \quad TBL   & 0.313* & 1.542 & 1.160 & 0.506 & 0.352 & 0.365 & 0.217 \\
    \quad LTY   & 0.933*** & 2.493 & 1.141 & 1.375** & 0.416** & 1.328 & 0.217 \\
    \quad LTR   & 0.471** & 1.831 & 0.772 & 0.624 & 0.359 & 0.514 & 0.393 \\
    \quad TMS   & 0.046 & 0.599 & 0.497 & -0.057 & 0.304 & -0.048 & 0.217 \\
    \quad DFY   & 0.077 & 1.044 & 0.862 & 0.000 & 0.308 & 0.000 & 0.217 \\
    \quad DFR   & -2.992 & -0.841 & -0.923 & -0.260 & 0.289 & 0.329 & -9.008 \\
    \quad INFL  & -0.809 & -0.171 & -0.751 & -0.950 & 0.236 & -2.056 & 1.451 \\
    \bottomrule
    \end{tabularx}
\end{table}
\clearpage

\begin{table}[htp!]
	\caption{Long-horizon Return Predictability}\label{Tab-Horizon}
			
    {\small\parindent=2em This table reports the 2 to 4-month horizon return predictability of market excess return predictors. Panel A provides the in- and out-of-sample return predictability results using the value-weighted aggregate joint news index ($\text{JointNews}^M$). Panel B shows the in-sample results for the 3- and 6-month bivariate predictive regression of monthly excess market returns on $\text{JointNews}^M$ and one of the other predictors, $Z_t$ (see Table \ref{Tab-InSample} for the full list of other predictors). ***, **, and * denote statistical significance at the 1\%, 5\%, and 10\% levels, respectively. The full sample period spans from July 1996 to December 2019, and the evaluation period for out-of-sample analysis is from July 2005 to December 2019.}
	\smallskip
	
	\renewcommand*{\arraystretch}{0.85}
	\small
	\centering
	\begin{tabularx}{\linewidth}{@{\extracolsep{\fill}}lS[table-format = -1.3, table-space-text-post ={***}]S[table-format = -1.3]S[table-format = 1.2]S[table-format = 1.3, table-space-text-post ={***}]S[table-format = 1.3, table-space-text-post ={***}]S[table-format = 1.3, table-space-text-post ={**}]}
	    \multicolumn{7}{@{\extracolsep{\fill}}l}{\textbf{Panel A: In- and Out-of-sample Analysis}} \\
        \toprule
        & \multicolumn{3}{c}{In-Sample} & \multicolumn{3}{c}{Out-of-Sample} \\
        \cmidrule{2-4}\cmidrule{5-7}
        Horizon & {$\widehat{\beta}$} & {$t$-stat.} & {$R^2$(\%)} & {$R_{OS}^2$(\%)} & {$\Delta$CER(\%)} & {SRatio} \\ 
        \midrule
        $h=2$   & -1.311*** & -3.197 & 4.43  & 4.919*** & 5.445*** & 0.792*** \\
        $h=3$   & -1.704*** & -3.168 & 4.99  & 4.815** & 5.436** & 0.836** \\
        $h=4$   & -1.686** & -2.354 & 3.52  & 1.881* & 0.783 & 0.391 \\
        \bottomrule
        \end{tabularx}
        \smallskip
        
        \renewcommand*{\arraystretch}{0.98}
        \setlength{\tabcolsep}{1pt}
        \begin{tabularx}{\linewidth}{@{\extracolsep{\fill}}lS[table-format = -1.3,table-space-text-post ={***}]S[table-format = -1.3,table-space-text-post ={***}]S[table-format = 1.2]S[table-format = -1.3,table-space-text-post ={***}]S[table-format = -1.3,table-space-text-post ={***}]S[table-format = 1.3]S[table-format = -1.3,table-space-text-post ={***}]S[table-format = -1.3,table-space-text-post ={***}]S[table-format = 1.3]} 
        \multicolumn{10}{@{\extracolsep{\fill}}l}{\textbf{Panel B: Horse Race}} \\
        \toprule
        & \multicolumn{3}{c}{$h = 2$} & \multicolumn{3}{c}{$h = 3$} & \multicolumn{3}{c}{$h = 4$} \\
        \cmidrule{2-4}\cmidrule{5-7}\cmidrule{8-10}
        Predictors & {$\widehat{\beta}$}  & {$\widehat{\phi}$}   & {$R^2$(\%)}  & {$\widehat{\beta}$}  & {$\widehat{\phi}$}   & {$R^2$(\%)}  & {$\widehat{\beta}$}  & {$\widehat{\phi}$}   & {$R^2$(\%)} \\
        \midrule
        \multicolumn{7}{@{\extracolsep{\fill}}l}{\textbf{Controlling for Self News}} \\
        $\text{SelfNews}^M$ & -1.379*** & -0.381 & 4.79  & -1.842*** & -0.763* & 5.96  & -1.785** & -0.536 & 3.87 \\
        \multicolumn{7}{@{\extracolsep{\fill}}l}{\textbf{Controlling for News Tone, Investor Sentiment and Attention}} \\
        News Tone & -1.269*** & -0.223 & 4.55  & -1.643*** & -0.324 & 5.17  & -1.616** & -0.363 & 3.68 \\
        Sent$^{BW}$ & -1.222*** & -0.899** & 6.53  & -1.566*** & -1.399** & 8.39  & -1.481** & -2.008*** & 8.63 \\
        Sent$^{PLS}$ & -1.193*** & -1.179*** & 8.03  & -1.541*** & -1.636*** & 9.63  & -1.468* & -2.150*** & 9.38 \\
        FEARS & -2.118*** & -1.214*** & 14.48 & -2.775*** & 0.538 & 12.26 & -2.974*** & 0.337** & 9.75 \\
        Attn$^{PLS}$ & -1.252*** & -1.076 & 6.99  & -1.720*** & -1.662 & 9.79  & -1.705** & -2.110 & 9.20 \\
        ASVI & -1.202*** & -0.495 & 6.61  & -1.248** & -1.482 & 9.53  & -0.875 & -2.372* & 10.40 \\
        RcrdHigh & -1.342*** & -0.610** & 5.39  & -1.746*** & -0.840*** & 6.21  & -1.734** & -1.086*** & 5.01 \\
        \multicolumn{7}{@{\extracolsep{\fill}}l}{\textbf{Controlling for Economic Predictors}} \\
        DP    & -1.259*** & 1.194* & 8.16  & -1.627*** & 1.794** & 10.62 & -1.584** & 2.478** & 11.37 \\
        DY    & -1.225*** & 1.161** & 7.94  & -1.572*** & 1.813** & 10.72 & -1.507** & 2.510** & 11.55 \\
        EP    & -1.301*** & 0.288 & 4.65  & -1.696*** & 0.254 & 5.11  & -1.679** & 0.225 & 3.58 \\
        DE    & -1.314*** & 0.319 & 4.70  & -1.710*** & 0.635 & 5.70  & -1.696** & 0.986 & 4.76 \\
        SVAR  & -1.204*** & 0.481 & 5.00  & -1.537*** & 0.753 & 5.94  & -1.504** & 0.816 & 4.33 \\
        BM    & -1.326*** & 0.743 & 5.87  & -1.730*** & 1.282* & 7.86  & -1.717*** & 1.900** & 8.12 \\
        NTIS  & -1.441*** & -0.995 & 6.98  & -1.907*** & -1.563 & 9.19  & -1.948** & -2.054* & 8.82 \\
        TBL   & -1.308*** & 0.350 & 4.75  & -1.699*** & 0.595 & 5.61  & -1.679** & 0.902 & 4.56 \\
        LTY   & -1.242*** & 0.453 & 4.94  & -1.599*** & 0.720 & 5.85  & -1.538** & 1.101 & 4.98 \\
        LTR   & -1.324*** & 0.279 & 4.63  & -1.711*** & 0.140 & 5.03  & -1.694** & 0.190 & 3.56 \\
        TMS   & -1.320*** & 0.058 & 4.44  & -1.728*** & 0.163 & 5.04  & -1.720** & 0.250 & 3.59 \\
        DFY   & -1.272*** & -0.292 & 4.65  & -1.681*** & -0.177 & 5.05  & -1.695** & 0.066 & 3.52 \\
        DFR   & -1.311*** & -0.026 & 4.43  & -1.700*** & 0.283 & 5.13  & -1.675** & 0.686 & 4.12 \\
        INFL  & -1.293*** & 0.146 & 4.48  & -1.776*** & -0.582 & 5.58  & -1.828** & -1.101** & 5.04 \\
        \multicolumn{7}{@{\extracolsep{\fill}}l}{\textbf{Controlling for IPO and M\&A Events}} \\
        IPO   & -1.312*** & 0.031 & 4.43  & -1.705*** & 0.001 & 4.99  & -1.696** & 0.417 & 3.74 \\
        MA$_n$ & -1.355*** & -0.271 & 4.61  & -1.730*** & -0.166 & 5.04  & -1.708** & -0.149 & 3.54 \\
        MA$_v$ & -1.336*** & -0.704** & 5.72  & -1.735*** & -0.888* & 6.37  & -1.735** & -1.233* & 5.45 \\
    \bottomrule
    \end{tabularx}
\end{table}
\clearpage

\begin{table}[htp!]
    \caption{Joint News Coverage and Mispricing}\label{Tab-Mispricing}
	
    {\small\parindent=2em This table reports differences in the degree of mispricing for monthly two-way sorted portfolios, sorted by joint news coverage of a firm and a key firm characteristic. The firm-level mispricing proxy (MISP) is constructed as in \cite{Stambaugh2015} by averaging each stock's rankings across 11 return anomalies. Each month, firms are first sorted into terciles based on one of four firm characteristics --- size, book-to-market ratio, idiosyncratic volatility, or analyst coverage. Within each tercile, firms are further grouped into high and low joint news coverage categories. The table reports MISP (H-L),  defined as the difference in MISP between the high and low joint news coverage portfolios. To ensure consistency, each firm's mispricing measure, joint news coverage, and firm characteristics are demeaned by their respective time-series averages. The $t$-statistics are reported in the brackets. ***, ** and * denote statistical significance at the 1\%, 5\%, and 10\% levels, respectively. The sample period spans from July 1996 to December 2019.}
    \smallskip
	
    \renewcommand*{\arraystretch}{1}
    \small
    \begin{tabularx}{\linewidth}{@{\extracolsep{\fill}}lcccc}
    \toprule
    & \multicolumn{4}{c}{Firm Characteristics} \\ \cmidrule{2-5}
    \multicolumn{1}{l}{JointNews} & 1     & 2     & 3     & Average \\
    \midrule
    & \multicolumn{4}{c}{\textit{Size}} \\ \cmidrule{2-5} 
    \multicolumn{1}{l}{MISP (H-L)} & 0.073 & 0.500*** & 0.282*** & 0.285*** \\
        & [0.70] & [4.22] & [3.05] & [3.88] \\ \cmidrule{2-5}
    & \multicolumn{4}{c}{\textit{BM}} \\ \cmidrule{2-5}
    \multicolumn{1}{l}{MISP (H-L)} & 0.192** & 0.182* & 0.040 & 0.138* \\
        & [1.98] & [1.89] & [0.31] & [1.76] \\ \cmidrule{2-5}
    & \multicolumn{4}{c}{\textit{IVOL}} \\ \cmidrule{2-5} \multicolumn{1}{l}{MISP (H-L)} & 0.307*** & 0.361*** & 0.134 & 0.267*** \\
        & [3.02] & [3.33] & [1.29] & [3.60] \\ \cmidrule{2-5}          
    & \multicolumn{4}{c}{\textit{Analyst}} \\ \cmidrule{2-5}
    \multicolumn{1}{l}{MISP (H-L)} & 0.459*** & 0.166 & 0.449*** & 0.358*** \\
          & [3.82] & [1.42] & [3.63] & [4.40] \\
    \bottomrule
    \end{tabularx}
\end{table}

\clearpage

\begin{table}[htp!]
    \caption{Market States and Return Predictability}\label{Tab-MktStates}
	
    {\small\parindent=2em This table provides in-sample estimation results from the predictive regression of monthly excess market returns on the aggregate joint news index, JointNews$^M$, across periods of high/low market uncertainty and frictions. We construct the market uncertainty indicators (PCA-Uncertainty and EW-Uncertainty) by aggregating several uncertainty measures, including VIX from CBOE, the economic uncertainty index (UNC), the treasury implied volatility index (TIV), the macro uncertainty (MU) and financial uncertainty (FU) index, the economic policy uncertainty index (EPU), and the disagreement index (DSA), through both PCA and equal-weighting. We also compute the market friction indicators (PCA-Friction and EW-Friction) using the equal-weighted short interest ratio (EWSI), value-weighted bid-ask spreads (BAS), and the price delay measure (DLY) in the same fashion. The overall market-state indicators (PCA-All and EW-All) use all the variables mentioned above, through both PCA and equal-weighting, respectively. Market states are classified as high (low) when the corresponding indicator falls in the top (bottom) 25\% of its distribution. The $t$-statistics reported in brackets are based on the Newey-West robust standard errors with 12 lags. ***, ** and * denote statistical significance at the 1\%, 5\%, and 10\% levels, respectively. The sample period spans from July 1996 to December 2018.}
    \smallskip
	
    \renewcommand*{\arraystretch}{1}
    \small
    \begin{tabularx}{\linewidth}{@{\extracolsep{\fill}}lcccccc}
	\toprule
    & \multicolumn{2}{c}{High} & \multicolumn{2}{c}{Low} & \multicolumn{2}{c}{Diff} \\
    \cmidrule{2-3}\cmidrule{4-5}\cmidrule{6-7}          
    Variables & {$\widehat{\beta}_H$}  & {$t$-stat.} & {$\widehat{\beta}_L$}  & {$t$-stat.} & {$\widehat{\beta}_H - \widehat{\beta}_L$}  & {$t$-stat.} \\
    \midrule
    \multicolumn{7}{@{\extracolsep{\fill}}l}{\textbf{Panel A. Market Uncertainty}}\\
    \quad PCA-Uncertainty & -1.010*** & [-2.72] & 0.044 & [0.13] & -1.054** & [-2.09] \\
    \quad EW-Uncertainty & -1.078*** & [-3.22] & 0.086 & [0.26] & -1.165** & [-2.46] \\
    \multicolumn{7}{@{\extracolsep{\fill}}l}{\textbf{Panel B. Market Frictions}}\\
    \quad PCA-Friction & -1.101*** & [-3.25] & -0.042 & [-0.19] & -1.060** & [-2.48] \\
    \quad EW-Friction & -1.137*** & [-3.20] & 0.031 & [0.14] & -1.168** & [-2.50] \\
    \multicolumn{7}{@{\extracolsep{\fill}}l}{\textbf{Panel C. Overall}}\\
    \quad PCA-All & -0.945** & [-2.52] & 0.515 & [1.35] & -1.460*** & [-2.73] \\
    \quad EW-All & -0.843*** & [-3.69] & 0.374 & [0.95] & -1.217** & [-2.48] \\
    \bottomrule
 \end{tabularx}
\end{table}

\clearpage

%% Appendices
\appendix
% \titleformat{\subsection}{\bf\large}{\thesubsection.}{1em}{}

\renewcommand{\thesection}{\Alph{section}}
\renewcommand{\theequation}{A\arabic{equation}}
\renewcommand{\thetable}{A\arabic{table}}
\renewcommand{\thefigure}{A\arabic{figure}}
\setcounter{equation}{0}
\setcounter{table}{0}
\setcounter{figure}{0}

\noindent \textbf{\LARGE Appendices}
\bigskip

Appendix \ref{App:Example} provides an illustrative example of joint news coverage. Appendix \ref{App:AddResults} summarizes the variables and notations used throughout the paper and presents additional results, including a replication of the main result in \cite{Ali2020} and an instrumental variable analysis of market return predictability.

\section{An Example of a Joint News Feed} \label{App:Example}
The following is an example of a joint news article extracted from Thomson Reuters News Archive (text data).  \bigskip

\includepdf[pages=-]{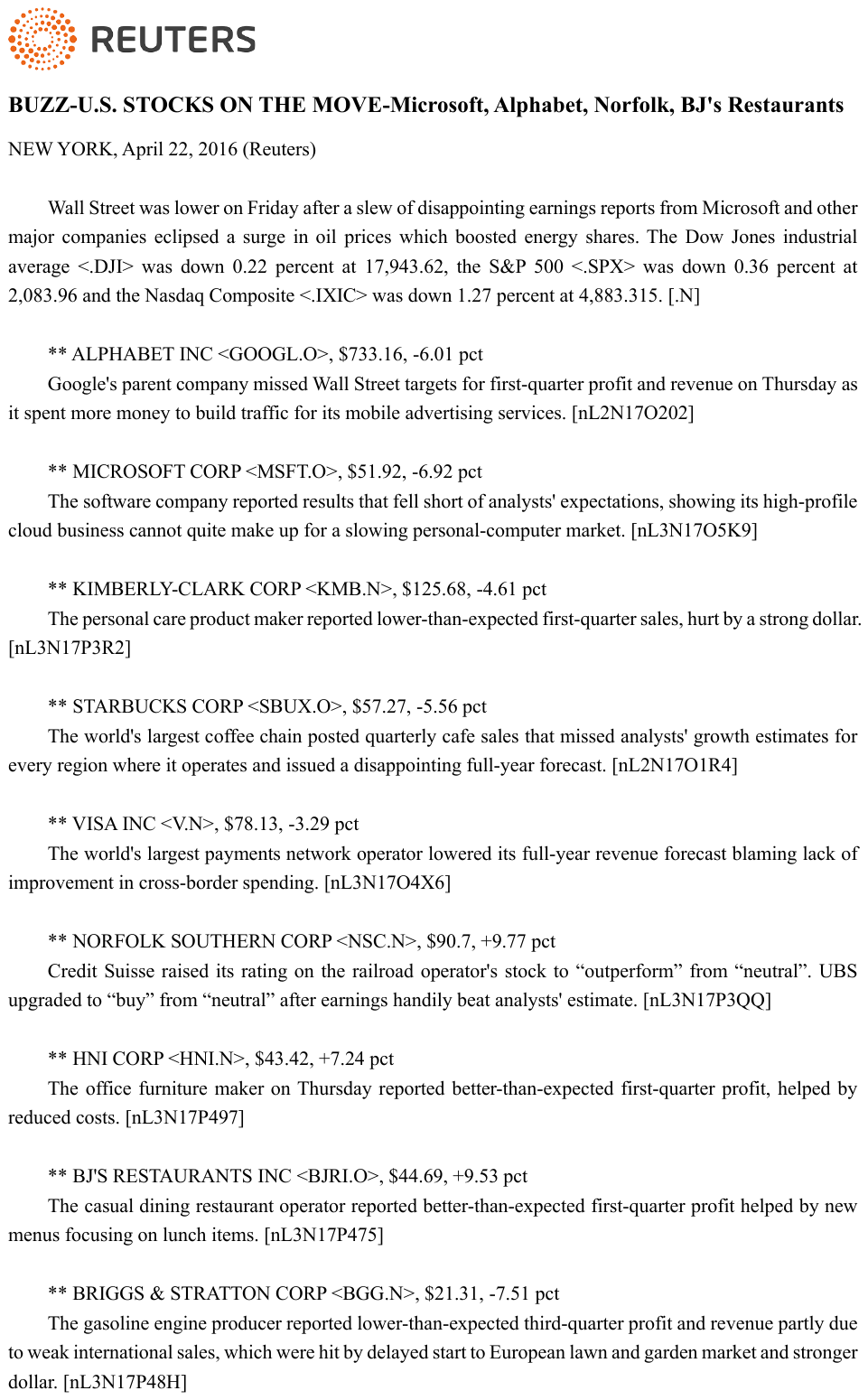}

\clearpage

\begin{figure}[htp!]
    \begin{center}
	\includegraphics[width = 0.95\textwidth]{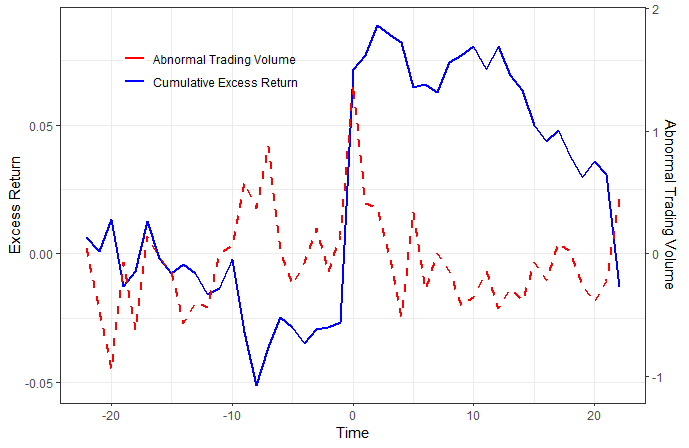}
    \end{center}
    \caption{\textbf{Cumulative Excess Returns and Abnormal Trading Volume of Norfork Southern Corp (NSC).} This figure plots the cumulative excess returns and abnormal trading volume of NSC over a $[-20, 20]$ day window surrounding the news ``Credit Suisse raised its rating on the railroad operator's stock to `outperform' from `neutral'. UBS upgraded to `buy' from `neutral' after earnings handily beat analysts' estimate." on April 22, 2016.} \label{Fig-App}
\end{figure}
\clearpage

\begin{table}[htp!]
    \caption{New IPs Searching Norfork Southern Corp (NSC) from Microsoft (MSFT)} \label{AppTab-IPs}
    \small
    \begin{tabularx}{\linewidth}{@{\extracolsep{\fill}}lccc}
    \toprule
    Date & \makecell{Number of NewIPs \\ from MSFT} & \makecell{Total Number \\ of NewIPs} & NewIPs \\
    \midrule
    2016/04/17 & 2     & 8     & \makecell[tl]{199.30.24.jdf, 207.46.13.eja} \smallskip\\
    2016/04/18 & 2     & 5     & \makecell[tl]{199.102.168.ceb, 157.55.2.abh} \smallskip\\
    2016/04/19 & 3     & 7     & \makecell[tl]{199.30.25.fhi, 157.55.2.fib, 192.189.187.jig} \smallskip\\
    2016/04/20 & 5     & 5     & \makecell[tl]{216.82.241.jdf, 199.30.24.jje, 65.90.142.fed,\\ 199.89.199.dbe, 64.233.172.bdf} \smallskip\\
    2016/04/21 & 4     & 12    & \makecell[tl]{199.67.131.bgd, 188.40.10.cfi, 136.243.64.cci,\\ 5.9.68.cdg} \smallskip\\
    \textbf{2016/04/22} & 23    & 42    & \makecell[tl]{208.71.189.ggf, 50.199.85.dci, 38.88.48.fdf,\\ 209.189.235.hjh, 151.194.64.jbj, 74.202.51.jcf,\\ 199.36.0.idd, 115.113.198.egd, 80.28.117.haf,\\ 208.24.125.gjj, 198.179.137.bgg, 199.94.1.ide,\\ 24.97.166.bic, 96.91.233.hji, 167.181.12.abj,\\ 157.55.12.geh, 66.169.194.aed,  64.15.16.ehg,\\ 209.220.148.abf, 170.74.231.jda, 208.28.3.jeb,\\ 192.68.178.jja, 52.87.220.adj} \smallskip\\
    2016/04/23 & 14    & 22    & \makecell[tl]{202.154.161.cih, 124.29.209.fgi, 66.162.2.gjc,\\ 199.30.24.abj, 184.189.250.aie, 71.97.49.gif,\\ 71.161.114.eff, 108.33.190.jcd, 46.4.101.cac,\\ 74.217.48.ide, 52.35.51.abb, 62.210.108.djb,\\ 173.203.18.hgj, 109.90.134.hji} \smallskip\\
    2016/04/24 & 2     & 6     & \makecell[tl]{199.30.24.jja, 159.203.75.bcj} \smallskip\\
    2016/04/25 & 11    & 33    & \makecell[tl]{217.9.192.gfj, 61.135.221.ajc, 65.127.84.hhg,\\ 206.208.93.edf, 91.219.236.gig, 72.1.86.aje,\\ 64.128.82.cah, 50.204.232.fjf, 125.227.5.cca,\\ 12.130.175.fed, 114.43.243.hbf} \smallskip\\
    2016/04/26 & 9     & 18    & \makecell[tl]{121.242.176.ahg, 208.5.220.bji, 101.226.33.caa,\\ 104.45.18.dad, 23.101.61.jca, 104.209.188.eaj,\\ 47.19.218.baj, 159.45.129.fcg, 38.97.88.fed} \smallskip\\
    2016/04/27 & 7     & 27    & \makecell[tl]{160.83.73.eag, 141.129.1.beg, 104.129.194.eca,\\ 199.30.16.dii, 173.52.51.jbd, 207.46.13.aef,\\ 131.253.26.ihd} \smallskip\\
    2016/04/28 & 6     & 9     & \makecell[tl]{206.41.50.gjc, 208.46.48.bdc, 65.55.210.bah,\\ 220.227.54.gge, 170.74.231.fed, 66.249.64.gai} \smallskip\\
    2016/04/29 & 7     & 13    & \makecell[tl]{158.82.194.eja, 190.113.30.cef, 65.55.210.icd,\\ 205.228.82.eef, 46.174.211.bic, 69.140.34.jdf,\\ 66.249.66.fbd} \smallskip\\
    2016/04/30 & 2     & 2     & \makecell[tl]{199.30.24.ihh, 81.200.55.cih} \smallskip\\
    2016/05/01 & 3     & 5     & \makecell[tl]{108.171.130.jje, 203.217.245.bbb, 66.102.6.hhj} \smallskip\\
    2016/05/02 & 2     & 7     & \makecell[tl]{192.48.47.jbj, 130.76.96.bfc} \\
    \bottomrule
    \end{tabularx}
\end{table}
\clearpage

\section{Variable Definitions and Additional Analysis} \label{App:AddResults}

\renewcommand{\theequation}{B\arabic{equation}}
\renewcommand{\thetable}{B\arabic{table}}
\renewcommand{\thefigure}{B\arabic{figure}}
\setcounter{equation}{0}
\setcounter{table}{0}
\setcounter{figure}{0}

\begin{table}[!htp]
    \caption{Variable Definitions} \label{Tab-VarDef}
   
    \renewcommand*{\arraystretch}{1}
    \small
    \begin{tabular}[t]{p{0.18\textwidth}p{0.80\textwidth}}
	\toprule
	Variable & Definition \\
	\midrule
	\multicolumn{2}{l}{\textbf{Returns}} \\
	\quad $R_m$ & Log return on the S\&P 500 Index in excess of the risk-free rate \\
	\quad $R_f$ & Risk-free rate, i.e., 1-month T-bill rate \medskip\\
		
	\multicolumn{2}{l}{\textbf{News-based Attention Measures}} \\
	\quad JointNews$_i$ & Firm-level average abnormal joint news coverage across all connected firms, weighted by the centrality of the connected firm. \\
	\quad JointNews$^M$ & Aggregate joint news index, value-weighted across firms \\
	\quad JointNews$^M_{ew}$ & Aggregate joint news index, equal-weighted across firms \\
	\quad SelfNews$_i$ & Firm-level abnormal self news coverage. \\
	\quad SelfNews$^M$ & Aggregate self news index, value-weighted across firms		\medskip\\
		
	\multicolumn{2}{l}{\textbf{News Tones, Investor Sentiment, and Composite Attention}} \\
	\quad \text{NewsTone}  & Negative news tone maintained by Thomson Reuters	\\
	\quad $\text{Sent}^\textit{BW}$ & Investor sentiment index estimated as the first principal component of six individual sentiment proxies \citep{Baker2006, Baker2007} \\
	\quad $\text{Sent}^\textit{PLS}$ & Investor sentiment index estimated by partial least squares \citep{Huang2015} \\
        \quad FEARS & Financial and Economic Attitudes Revealed by Search (FEARS) index \citep{DaEngelbergGao2015} \\
	\quad $\text{Attn}^\textit{PLS}$ & Investor attention index estimated by partial least squares \citep{Chen2020} \\
	\quad ASVI & Abnormal Google search volume for major stock indices \citep{LiuPengTang2023} \\
        \quad RcrdHigh & Nearness to the Dow historical high \citep{Chen2020} \medskip\\
        
        \multicolumn{2}{l}{\textbf{Economic Predictors \citep{Goyal2008}}} \\
	\quad DP & Log dividend-price ratio \\
	\quad DY & Log dividend-yield ratio\\
	\quad EP & Log earnings-price ratio \\
	\quad DE & Log dividend buyout ratio \\
	\quad SVAR & Stock return variance \\
	\quad BM & Book-to-market ratio \\
	\quad NTIS & Net equity expansion \\
	\quad TBL & Treasury bill rate \\
	\quad LTY & Long-term bond yield \\
	\quad LTR & Long-term bond return \\
	\quad TMS & Term spread \\
	\quad DFY & Default yield spread \\
	\quad DFR & Default return spread \\
	\quad INFL & Inflation rate \\
	\bottomrule
    \end{tabular}
\end{table}
\clearpage

\begin{table}[htp!]
    \caption*{Table \ref{Tab-VarDef} (Cont'd) \quad Variable Definitions}	
    \small
    \begin{tabular}[t]{p{0.18\textwidth}p{0.80\textwidth}}
	\toprule
	Variable & Definition  \\
	\midrule
        \multicolumn{2}{l}{\textbf{IPO and M\&A Events}} \\
        \quad IPO & monthly number of IPO cases, maintained by Jay Ritter\\
        \quad MA$_n$ & monthly number of M\&A cases, maintained by IMAA Institute\\
        \quad MA$_v$ & monthly gross value of M\&A cases, maintained by IMAA Institute \medskip\\  
        
	\multicolumn{2}{l}{\textbf{Market Uncertainty and Frictions}} \\
	\quad VIX &  CBOE volatility index\\
	\quad UNC & Economic uncertainty index \citep{BaliBrownCaglayan2014} \\
	\quad TIV & Treasury implied volatility \citep{ChoiMuellerVedolin2017}\\
	\quad MU & Macro uncertainty index \citep{JuradoLudvigsonNg2015} \\
	\quad FU & Financial uncertainty index \citep{JuradoLudvigsonNg2015} \\
	\quad EPU & Economic policy uncertainty index \citep{BakerBloomDavis2016} \\
	\quad DSA & Disagreement index \citep{Huang2021} \\
	\quad EWSI & Equal-weighted short-interest ratio \citep{RapachRinggenbergZhou2016} \\
	\quad BAS & Value-weighted average of bid-ask spreads across S\&P500 firms\\
	\quad DLY & Value-weighted average of the second type of price delay measure \citep{Hou2005} across S\&P500 firms \medskip\\

	\multicolumn{2}{l}{\textbf{Variables for Cross-Sectional Analysis}} \\
	\quad ASV & Abnormal Google search volume. The log of SVI during the month minus the log of average SVI during the previous 12 months, skipping the most recent month\\
	\quad New IP & Number of new IPs in EDGAR search traffic, defined as the unique IP address that has not searched the focal stock in the past 18 months \\
	\quad New IP$_c$ & Number of new IPs from connected stocks in EDGAR search traffic, defined as the unique IP address that has not searched the focal stock in the past 18 months but searched the connected stocks in the past 18 months \\
	%\quad New IP$_o$ & Number of new IPs from other stocks in EDGAR search traffic, defined as New IP net of New IP$_c$ \\
	%\quad Exist IP & Number of unique IP addresses that have searched the focal stock in the past 18 months \\
        \quad ln(Size) & Log market cap \\
        \quad ln(BM) & Log book-to-market ratio \\
	\quad IVOL & Idiosyncratic volatility based on the Fama-French 3 factor model \\
	% \quad MOM & Average returns from $t-2$ to $t-12$ \\
        \quad AbnTnvr & Standardized abnormal turnover as in \citet{Chordia2007} \\
        % \quad Abn Ret & Characteristic-adjusted return as in \citet{Daniel1997} \\
        \quad AdExp/Sales & Ratio between advertisement expense and sales in the previous fiscal year, where we set advertisement expenditure to zero if it is missing in Compustat \\
        \quad ln(1+Analyst) & Log number of analysts in I/B/E/S  \\
	\bottomrule
    \end{tabular}	
\end{table}%
\clearpage

\begin{landscape}

\begin{table}[htp!]
    \caption{Correlation Coefficients between Joint News and Other Predictors}\label{Tab-Corr}	
	
    {\small\parindent=2em This table reports simple correlation coefficients (in \%) between all the time series variables. The variable definitions are given in Table \ref{Tab-VarDef}. The sample period is from July 1996 to December 2019.}
    \smallskip
	
    \small
    \renewcommand{\arraystretch}{0.9}
    \setlength{\tabcolsep}{2pt}
    \begin{tabularx}{\linewidth}{@{\extracolsep{\fill}}lcccccccccccccccccccccccccc}
    \toprule
    & 1     & 2     & 3     & 4     & 5     & 6     & 7     & 8     & 9     & 10    & 11    & 12    & 13    & 14    & 15    & 16    & 17    & 18    & 19    & 20    & 21    & 22    & 23    & 24    & 25    & 26        \\
    \midrule
    JointNews$^{M}$ &       &       &       &       &       &       &       &       &       &       &       &       &       &       &       &       &       &       &       &       &       &       &       &       &       &  \\
    JointNews$_{ew}^{M}$ & 90    &       &       &       &       &       &       &       &       &       &       &       &       &       &       &       &       &       &       &       &       &       &       &       &       &  \\
    SelfNews$^{M}$ & -19   & -16   &       &       &       &       &       &       &       &       &       &       &       &       &       &       &       &       &       &       &       &       &       &       &       &  \\
    News Tone & 19    & 12    & -4    &       &       &       &       &       &       &       &       &       &       &       &       &       &       &       &       &       &       &       &       &       &       &  \\
    Sent$^{BW}$ & 9     & 7     & -2    & 25    &       &       &       &       &       &       &       &       &       &       &       &       &       &       &       &       &       &       &       &       &       &  \\
    Sent$^{PLS}$ & 9     & 7     & -5    & 37    & 69    &       &       &       &       &       &       &       &       &       &       &       &       &       &       &       &       &       &       &       &       &  \\
    FEARS & 4     & 1     & -24   & -2    & 8     & -9    &       &       &       &       &       &       &       &       &       &       &       &       &       &       &       &       &       &       &       &  \\
    Attn$^{PLS}$ & 4     & 0     & 2     & 6     & 54    & 39    & -11   &       &       &       &       &       &       &       &       &       &       &       &       &       &       &       &       &       &       &  \\
    ASVI & 24    & 17    & 9     & 25    & 18    & 3     & -12   & 28    &       &       &       &       &       &       &       &       &       &       &       &       &       &       &       &       &       &  \\
    RcrdHigh & -7    & -1    & -2    & -49   & 28    & -20   & 6     & 26    & -2    &       &       &       &       &       &       &       &       &       &       &       &       &       &       &       &       &  \\
    DP    & -5    & 0     & -4    & 3     & -51   & -38   & -4    & -42   & 4     & -35   &       &       &       &       &       &       &       &       &       &       &       &       &       &       &       &  \\
    DY    & -8    & -3    & -5    & -3    & -54   & -42   & -6    & -42   & -9    & -30   & 97    &       &       &       &       &       &       &       &       &       &       &       &       &       &       &  \\
    EP    & -4    & -2    & -2    & -42   & -3    & -32   & 4     & -6    & 14    & 51    & -2    & -2    &       &       &       &       &       &       &       &       &       &       &       &       &       &  \\
    DE    & 0     & 1     & -1    & 37    & -23   & 9     & -5    & -15   & -11   & -62   & 48    & 48    & -88   &       &       &       &       &       &       &       &       &       &       &       &       &  \\
    SVAR  & -23   & -12   & -6    & -44   & 4     & -15   & 18    & -11   & -63   & 49    & -28   & -20   & 27    & -38   &       &       &       &       &       &       &       &       &       &       &       &  \\
    BM    & 2     & 8     & -5    & -17   & -57   & -46   & 6     & -61   & 3     & -20   & 69    & 67    & 40    & -3    & -7    &       &       &       &       &       &       &       &       &       &       &  \\
    NTIS  & -14   & -15   & -1    & -4    & -13   & -21   & -5    & -6    & 10    & 1     & 50    & 48    & -6    & 28    & -22   & 22    &       &       &       &       &       &       &       &       &       &  \\
    TBL   & -2    & 4     & -10   & -1    & -59   & -24   & -4    & -68   & -14   & -31   & 62    & 63    & 3     & 26    & -6    & 68    & 26    &       &       &       &       &       &       &       &       &  \\
    LTY   & -17   & -10   & -10   & -15   & -43   & -38   & 6     & -58   & -16   & 9     & 57    & 58    & 23    & 6     & 7     & 62    & 54    & 75    &       &       &       &       &       &       &       &  \\
    LTR   & 5     & 3     & 11    & 17    & 5     & 3     & -8    & -3    & 15    & -11   & 4     & -1    & 5     & -3    & -19   & 4     & -6    & -4    & 4     &       &       &       &       &       &       &  \\
    TMS   & 16    & 17    & -5    & 15    & -43   & 4     & -7    & -49   & -6    & -58   & 32    & 31    & -21   & 33    & -17   & 35    & -19   & 69    & 4     & -10   &       &       &       &       &       &  \\
    DFY   & 13    & 7     & -3    & 43    & -33   & -1    & -3    & -23   & 4     & -70   & 60    & 57    & -50   & 71    & -59   & 33    & 45    & 38    & 21    & 2     & 35    &       &       &       &       &  \\
    DFR   & -3    & -1    & -13   & -4    & -6    & 1     & -2    & -5    & -42   & 0     & 1     & 10    & -19   & 16    & 24    & -4    & -1    & 8     & 4     & -48   & 8     & 11    &       &       &       &  \\
    INFL  & -12   & -9    & 8     & -19   & 7     & 0     & -26   & -2    & -10   & 9     & -17   & -16   & 6     & -14   & 22    & -8    & -3    & -13   & -14   & -9    & -5    & -26   & -7    &       &       &  \\
    IPO   & 2     & 3     & 9     & -33   & 19    & 1     & 0     & 45    & -11   & 41    & -39   & -36   & 13    & -31   & 22    & -38   & -44   & -56   & -50   & 0     & -30   & -49   & -6    & 13    &       &  \\
    MA$_n$ & -19   & -20   & -13   & -33   & 7     & -24   & 10    & 33    & 7     & 58    & -15   & -13   & 18    & -24   & 18    & -24   & 22    & -34   & 16    & -2    & -70   & -39   & -5    & 2     & 27    &  \\
    MA$_v$ & -5    & -5    & -6    & -37   & 3     & -17   & 9     & 7     & 0     & 45    & -17   & -16   & 14    & -21   & 16    & -9    & 14    & -17   & 14    & -7    & -42   & -26   & -9    & 9     & 19    & 52  \\
    \bottomrule	
    \end{tabularx}
\end{table}

\end{landscape}
\clearpage

\begin{table}[htp!]
    \caption{Correlation Coefficients between Joint News and Topic-specific News Measures}
    \label{Tab-CorrTopic}
    \small
    \setlength{\tabcolsep}{9pt}
    \begin{tabularx}{\textwidth}{@{\extracolsep{\fill}}lcccccccccccc}
	
    \multicolumn{13}{@{\extracolsep{\fill}}l}{\textbf{Panel A: Topic-specific Sentiment, Frequency, and Entropy \citep{Calomiris2019} }} \\
    \toprule
    & 1     & 2     & 3     & 4     & 5     & 6     & 7     & 8     & 9     & 10    & 11    & 12   \\
    \midrule
    $\text{JointNews}^M$ &       &       &       &       &       &       &       &       &       &       &       &  \\
    Entropy & 17    &       &       &       &       &       &       &       &       &       &       &  \\
    ArtCount & 5     & -55   &       &       &       &       &       &       &       &       &       &  \\
    sMkt  & -21   & -23   & -12   &       &       &       &       &       &       &       &       &  \\
    fMkt  & 11    & 40    & 24    & -23   &       &       &       &       &       &       &       &  \\
    sComms & -2    & 45    & 9     & 5     & 61    &       &       &       &       &       &       &  \\
    fComms & -15   & -60   & -6    & 36    & -59   & -59   &       &       &       &       &       &  \\
    sCorp & -12   & 17    & -5    & 66    & 37    & 57    & -15   &       &       &       &       &  \\
    fCorp & 8     & -36   & 15    & 13    & -45   & -34   & 24    & -32   &       &       &       &  \\
    sCredit & -24   & -33   & -16   & 83    & -16   & -8    & 49    & 54    & 6     &       &       &  \\
    fCredit & -12   & -67   & 61    & 8     & -14   & -13   & 11    & 6     & -7    & 1     &       &  \\
    sGovt & -3    & -49   & 69    & 29    & 29    & 26    & 4     & 38    & 25    & 23    & 53    &  \\
    fGovt & 0     & 67    & -77   & -10   & -23   & 3     & -15   & -10   & -31   & -11   & -67   & -86 \\
    \bottomrule	
    \end{tabularx}
    \medskip

    \begin{tabularx}{\linewidth}{@{\extracolsep{\fill}}lccccc}	
    \multicolumn{6}{@{\extracolsep{\fill}}l}{\textbf{Panel B: Topic Attention \citep{Bybee2021} }} \\
    \toprule
    & 1     & 2     & 3     & 4     & 5     \\
    \midrule
    $\text{JointNews}^M$    &       &       &       &       &  \\
    Recession & 1     &       &       &       &  \\
    Problems & 9     & 37    &       &       &  \\
    Record High & -12   & -5    & -24   &       &  \\
    Option/VIX & 8     & 6     & -2    & -18   &  \\
    Convertible/preferred & 9     & -19   & -4    & -9    & -4 \\
    \bottomrule	
    \end{tabularx}
    \medskip 

    \setlength{\tabcolsep}{3pt}
    \begin{tabularx}{\linewidth}{@{\extracolsep{\fill}}lccccccccccccccccc}
    \multicolumn{18}{@{\extracolsep{\fill}}l}{\textbf{Panel C: Topic-specific Joint News}} \\
    \toprule
    & 1     & 2     & 3     & 4     & 5  & 6     & 7     & 8     & 9     & 10 & 11     & 12     & 13     & 14     & 15  & 16  & 17 \\
    \midrule
    JointNews$^M$ &       &       &       &       &       &       &       &       &       &       &       &       &       &       &       &       &  \\
    JointNews$_{ew}^M$ & 90    &       &       &       &       &       &       &       &       &       &       &       &       &       &       &       &  \\
    JointNews$^{\textit{sply}}$ & -11   & -11   &       &       &       &       &       &       &       &       &       &       &       &       &       &       &  \\
    JointNews$^{\textit{ptnr}}$ & -25   & -25   & 38    &       &       &       &       &       &       &       &       &       &       &       &       &       &  \\
    JointNews$^{\textit{ma}}$ & -31   & -28   & 47    & 72    &       &       &       &       &       &       &       &       &       &       &       &       &  \\
    JointNews$^{\textit{prnt}}$ & -9    & -10   & 3     & 10    & 7     &       &       &       &       &       &       &       &       &       &       &       &  \\
    JointNews$^{\textit{legal}}$ & 9     & 6     & 13    & -9    & -11   & -12   &       &       &       &       &       &       &       &       &       &       &  \\
    JointNews$^{\textit{regl}}$ & 6     & 6     & 20    & 17    & 21    & -2    & 2     &       &       &       &       &       &       &       &       &       &  \\
    JointNews$^{\textit{lbr}}$ & 20    & 16    & -12   & -9    & -22   & -7    & 1     & 5     &       &       &       &       &       &       &       &       &  \\
    JointNews$^{\textit{cgvn}}$ & 1     & 4     & -4    & 3     & -7    & 12    & -4    & -5    & -2    &       &       &       &       &       &       &       &  \\
    JointNews$^{\textit{mgmt}}$ & -6    & -6    & 3     & -3    & 1     & 11    & 1     & 2     & -2    & 1     &       &       &       &       &       &       &  \\
    JointNews$^{\textit{cstm}}$ & -9    & -9    & 18    & 18    & 24    & -2    & 3     & 4     & -7    & -2    & -3    &       &       &       &       &       &  \\
    JointNews$^{\textit{ivst}}$ & 1     & 4     & 7     & -2    & 7     & 1     & -3    & 2     & -4    & -2    & 5     & 9     &       &       &       &       &  \\
    JointNews$^{\textit{nres}}$ & -17   & -14   & 45    & 57    & 72    & 3     & -3    & 38    & -21   & -5    & 1     & 19    & 4     &       &       &       &  \\
    JointNews$^{\textit{engy}}$ & -23   & -20   & 59    & 60    & 78    & 4     & 0     & 35    & -20   & -6    & 10    & 24    & 7     & 86    &       &       &  \\
    JointNews$^{\textit{tech}}$ & 4     & 6     & 28    & 32    & 31    & 7     & -4    & 30    & 5     & -1    & 6     & 1     & 2     & 38    & 41    &       &  \\
    JointNews$^{\textit{gptc}}$ & 15    & 10    & 1     & -12   & -13   & -6    & 30    & 26    & 4     & 5     & 4     & 4     & -2    & 6     & 3     & 9     &  \\
    JointNews$^{\textit{resid}}$ & 42    & 35    & -14   & -12   & -17   & -10   & -8    & -9    & 7     & 4     & -10   & -6    & 5     & -13   & -17   & -10   & -7 \\
    \bottomrule	
    \end{tabularx}
\end{table}
\clearpage

\begin{table}[htp!]
    \caption{Forecasting Firm-Level Returns Controlling for Connected-Firm Momentum Spillovers using CRSP Stocks} \label{AppTab-CFRet}
    {\small\parindent=2em This table reports the Fama-MacBeth regression results of next-month stock returns (in bps) on connected-firm momentum spillovers. Specifically, we follow \cite{Ali2020} to compute the connected-firm portfolio returns (CFRet) based on firm links identified through shared analyst coverage (ana), customer-supplier (cus), common industry (ind), technology (tech), or geographic (geo) links. The $t$-statistics reported in brackets are based on the Newey-West robust standard errors with six lags. ***, **, and * denote statistical significance at the 1\%, 5\%, and 10\% levels, respectively. The intercept is suppressed. The sample period spans from July 1996 to December 2019.}
    \smallskip
	
    \small
    \renewcommand*{\arraystretch}{1}
    \setlength{\tabcolsep}{2pt}
    \begin{tabularx}{\linewidth}{@{\extracolsep{\fill}}lcccccc}
		\toprule
        Variables & {(1)}   & {(2)}   &  {(3)} &  {(4)} & {(5)}   & {(6)} \\
        \midrule
        $\text{CFRet}^{ana}$ & 0.036*** &       &       &       &       & 0.038** \\
              & [4.11] &       &       &       &       & [2.26] \\
        $\text{CFRet}^{cus}$ &       & 0.028*** &       &       &       & 0.005 \\
              &       & [3.87] &       &       &       & [0.40] \\
        $\text{CFRet}^{ind}$ &       &       & 0.080*** &       &       & -0.001 \\
              &       &       & [4.79] &       &       & [-0.02] \\
        $\text{CFRet}^{tech}$ &       &       &       & 0.097*** &       & -0.003 \\
              &       &       &       & [3.17] &       & [-0.06] \\
        $\text{CFRet}^{geo}$ &       &       &       &       & 0.011* & 0.019 \\
              &       &       &       &       & [1.80] & [1.05] \\
        Controls  & No    & No    & No    & No    & No    & No \\
        Observations & 472,794  & 260,423  & 1,389,936  & 510,172  & 1,161,962  & 36,953  \\
        Adj.~$R^2$ (\%) & 0.8 & 0.4 & 0.6 & 0.7 & 0.1 & 6.3 \\
    \bottomrule
    \end{tabularx}
\end{table}
\clearpage

\begin{table}[htp!]
    \caption{Instrumental Variable Analysis}\label{Tab-IV}
    
    {\small\parindent=2em This table reports results from the instrumental variable analysis of market return predictability. $\text{JointNews}^M$ denotes the value-weighted aggregate joint news index. $I_\text{Dist}$ is a ``distraction" indicator that equals one if the monthly \citet{Eisensee2007} news pressure falls within the top tercile of its annual distribution. The $t$-statistics reported in the brackets are based on the Newey-West robust standard errors with optimal lags. ***, ** and * denote statistical significance at the 1\%, 5\%, and 10\% levels, respectively. The sample period spans from July 1996 to December 2019.}
    \smallskip
    
    \renewcommand*{\arraystretch}{1}
    \small
    \centering
    \begin{tabularx}{\linewidth}{@{\extracolsep{\fill}}lcccc}
	
    \multicolumn{5}{@{\extracolsep{\fill}}l}{\textbf{Panel A: Univariate Contrast}} \\
    \toprule
    Variables & {Total} & {$I_\text{Dist} = 0$} & {$I_\text{Dist} = 1$} & {Diff.} \\
    \midrule
    JointNews$^M$ (avg.) & 0.000 & 0.061 & -0.242* & 0.303** \\
     & [0.00] & [0.80] & [-1.94] & [2.06] \\
    Observations & 282   & 225   & 57    &  \\
   \bottomrule
   \\
    
    \multicolumn{5}{@{\extracolsep{\fill}}l}{\textbf{Panel B: Two-Stage Least Squares}} \\ 
    \toprule
    & \multicolumn{2}{c}{First Stage (JointNews$^M$)} & \multicolumn{2}{c}{Second Stage ($R_{m, t+1}$)} \\ 
    \cmidrule{2-3}\cmidrule{4-5} 
    Variables & $\beta$  & $t$-stat & $\beta$  & $t$-stat \\
    \midrule
    Predicted JointNews$^M$ &       &       & $-$3.474** & [$-$2.08] \\
    $I_\text{Dist}$ & $-$0.308** & [$-$2.31] &       &  \\ 
    \midrule
    Observations & \multicolumn{2}{c}{281}   &  \multicolumn{2}{c}{281}\\
    First-stage $F$-statistic & \multicolumn{2}{c}{5.45**}  & \multicolumn{2}{c}{---} \\
    Anderson-Rubin Weak-IV-Robust Test & \multicolumn{2}{c}{---} & \multicolumn{2}{c}{5.78**} \\
    Adj. $R^2$ (\%) & \multicolumn{2}{c}{1.19} & \multicolumn{2}{c}{0.65} \\
    \bottomrule
    \end{tabularx}
\end{table}

\end{document}